\documentstyle[epsf,rotating]{mn}

\newif\ifAMStwofonts

\ifoldfss
  \ifCUPmtlplainloaded \else
    \NewTextAlphabet{textbfit} {cmbxti10} {}
    \NewTextAlphabet{textbfss} {cmssbx10} {}
    \NewMathAlphabet{mathbfit} {cmbxti10} {} 
    \NewMathAlphabet{mathbfss} {cmssbx10} {} 
  \fi
  \ifAMStwofonts
    \ifCUPmtlplainloaded \else
      \NewSymbolFont{upmath} {eurm10}
      \NewSymbolFont{AMSa} {msam10}
      \NewMathSymbol{\upi}     {0}{upmath}{19}
      \NewMathSymbol{\umu}     {0}{upmath}{16}
      \NewMathSymbol{\upartial}{0}{upmath}{40}
      \NewMathSymbol{\leqslant}{3}{AMSa}{36}
      \NewMathSymbol{\geqslant}{3}{AMSa}{3E}

      \let\leq=\leqslant 
      \let\geq=\geqslant 
    \fi
  \fi
\fi 

\ifnfssone
  \newmathalphabet{\mathit}
  \addtoversion{normal}{\mathit}{cmr}{m}{it}
  \addtoversion{bold}{\mathit}{cmr}{bx}{it}
  \newmathalphabet{\mathbfit} 
  \addtoversion{normal}{\mathbfit}{cmr}{bx}{it}
  \addtoversion{bold}{\mathbfit}{cmr}{bx}{it}
  \newmathalphabet{\mathbfss} 
  \addtoversion{normal}{\mathbfss}{cmss}{bx}{n}
  \addtoversion{bold}{\mathbfss}{cmss}{bx}{n}
  \ifAMStwofonts
    \ifCUPmtlplainloaded \else
      \UseAMStwoboldmath
      \makeatletter
      \new@mathgroup\upmath@group
      \define@mathgroup\mv@normal\upmath@group{eur}{m}{n}
      \define@mathgroup\mv@bold\upmath@group{eur}{b}{n}
      \edef\UPM{\hexnumber\upmath@group}
      \new@mathgroup\amsa@group
      \define@mathgroup\mv@normal\amsa@group{msa}{m}{n}
      \define@mathgroup\mv@bold\amsa@group{msa}{m}{n}
      \edef\AMSa{\hexnumber\amsa@group}
      \makeatother
      \mathchardef\upi="0\UPM19
      \mathchardef\umu="0\UPM16
      \mathchardef\upartial="0\UPM40
      \mathchardef\leqslant="3\AMSa36
      \mathchardef\geqslant="3\AMSa3E

      \let\leq=\leqslant 
      \let\geq=\geqslant 
    \fi
  \fi
\fi 

\ifnfsstwo
  \DeclareMathAlphabet{\mathbfit}{OT1}{cmr}{bx}{it}
  \SetMathAlphabet\mathbfit{bold}{OT1}{cmr}{bx}{it}
  \DeclareMathAlphabet{\mathbfss}{OT1}{cmss}{bx}{n}
  \SetMathAlphabet\mathbfss{bold}{OT1}{cmss}{bx}{n}
  \ifAMStwofonts
    \ifCUPmtlplainloaded \else
      \DeclareSymbolFont{UPM}{U}{eur}{m}{n}
      \SetSymbolFont{UPM}{bold}{U}{eur}{b}{n}
      \DeclareSymbolFont{AMSa}{U}{msa}{m}{n}
      \DeclareMathSymbol{\upi}{0}{UPM}{"19}
      \DeclareMathSymbol{\umu}{0}{UPM}{"16}
      \DeclareMathSymbol{\upartial}{0}{UPM}{"40}
      \DeclareMathSymbol{\leqslant}{3}{AMSa}{"36}
      \DeclareMathSymbol{\geqslant}{3}{AMSa}{"3E}

      \let\leq=\leqslant 
      \let\geq=\geqslant 
    \fi
  \fi
\fi 

\ifCUPmtlplainloaded \else
  \ifAMStwofonts \else 
    \def\upi{\pi}
    \def\umu{\mu}
    \def\upartial{\partial}
  \fi
\fi

\title{The field of NGC 6397 as a test for Galactic models}
\author[V. Castellani et al.]
       {V. Castellani$^{1,2}$, S. Degl'Innocenti $^{1,2}$, S. Petroni$^1$, G. Piotto$^3$\\
  $^1$ Dipartimento di Fisica, Universit\'a di Pisa, Piazza Torricelli 2, 56126 Pisa, Italy \\
  $^2$ Istituto Nazionale di Fisica Nucleare, Sezione di Pisa, via Livornese 1291, S.Piero a Grado, 56100 Pisa\\
  $^3$ Dipartimento di Astronomia, Universit\'a di Padova, Vicolo dell'Osservatorio 5, 35122 Padova, Italy
}

\date{}

\pagerange{\pageref{firstpage}--\pageref{lastpage}}
\pubyear{2000}

\newcommand{\gcu}{\mbox{$\, \stackrel{ > }{ _{\sim} } \,$}}
\newcommand{\lcu}{\mbox{$\, \stackrel{ < }{ _{\sim} } \,$}}

\begin{document}

\maketitle

\label{firstpage}

\begin{abstract}

Taking advantage of recent HST data for field stars in the region of
the Galactic globular cluster NGC~6397, we tested the predictions of
several Galactic models with star counts reaching a largely unexplored
range of magnitudes, down to $V$$\sim$ 26.5.  After updating the input
stellar ($V-I$) colors, we found that the two-component
Bahcall-Soneira (B\&S) model can be put in satisfactory agreement with
observations for suitable choices of disk/spheroid luminosity
functions.  However if one assumes the Gould, Bahcall and Flynn (1996,
1997) disk luminosity function (LF) together with the Gould, Flynn and
Bahcall (1998) spheroid LF, there is no way to reconcile the predicted
and observed $V\,$-magnitude distribution.  We also analysed the
agreement between observed and predicted magnitude and color
distributions for two selected models with a thick disk component.
Even in this case there are suitable combinations of model parameters
and faint magnitude LFs which can give a reasonable agreement with
observational star counts both in magnitude and in color, the above
quoted combination of Gould et al. (1997, 1998) LFs giving again
predictions in clear disagreement with observations.

\end{abstract}


\section{Introduction}
 
Since the very beginning of modern astronomy, the distribution of
stars over the night sky has been understood as evidence for the
spatial distribution we now know as ``the Galaxy". Therefore, the use
of star counts to constrain the Galactic structure has been the goal
of generations of astronomers. In the long run, relevant progresses on
that matter has been allowed by the increased amount of observational
data, by the availability of modern computers and, last but not least,
by the original approach to the problem presented twenty years ago in
a seminal paper by Bahcall \& Soneira (1980, see also 1984).  Since
that time star counts began an effective way of investigating the
broad properties of stellar populations in our Galaxy (see e.g.
Gilmore 1981, 1984, Gilmore \& Reid 1983 and for a review Bahcall 1986
and Gilmore, Wyse \& Kuijken 1989).

Galactic models, which predicts star counts in selected areas of the
sky and for selected intervals of magnitude, give, at the same time,
constraints about such relevant issues as the amount of mass in the
form of stars and, in particular, about the contribution of faint
stars to the dark-matter problem. One naturally expects that the
contribution of intrinsically faint stars increases as the limiting
magnitude of star counts is increased. Thus the limiting magnitude of
the observational samples is a critic factor governing the possibility
of constraining the Galactic abundance of such kind of stars. However,
the difficulty in distinguishing faint images of stars from galaxies
has limited most of the ground based work to V $\lcu$ 20, reaching as
extreme limit V$\approx$ 22 (e.g. Bahcall 1986, Reid \& Majewski 1993,
Haywood et al. 1997 and references therein).

This situation has been recently improved by observational data from
the Hubble Space Telescope (HST). HST observations reach very faint
magnitudes (V$\approx$30, see e.g. Williams et al. 1996) and even
though the number of stars in the field of view is generally quite
small they can be used to test Galactic models at the best resolution
available. However the tests of Galactic models at faint magnitudes
available in the literature are still very few and mainly based on a
restricted number of stars (see Mendez et al. 1996, Reid et al. 1996,
Basilio et al. 1996, Mendez \& Guzman 1998).

\begin{figure}
\label{CMD}
\centerline{\epsfxsize= 7 cm \epsfbox{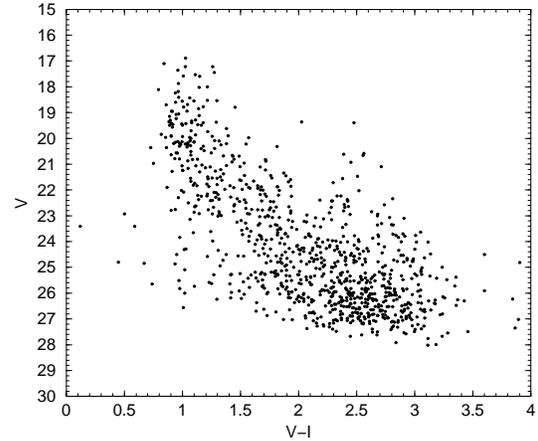}}
\caption{The $(V,V-I)$ CM diagram for the sample of field stars studied.}
\end{figure}

In this paper we will take advantage of recent HST observations of the
Galactic globular cluster NGC~6397, to discuss a rich sample of about
thousand bona fide Galactic field stars which should be fairly
complete (more than 95\% detection level) down to $V \sim$ 25, with a
reliable evaluation of the completeness reaching $V \sim$ 27 (King et
al. 1998, hereafter KACP). This sample obviously gives the exciting
opportunity of testing Galactic models over an almost unexplored range
of magnitudes, with possible relevant outcomes concerning the
abundance of faint stars.  For the analysis we will use observational
data of field stars in the direction of NGC 6397 from the second epoch
HST observations.  Field stars have been obtained by an accurate
separation from cluster stars based on proper motions analysis (KACP).
The area covered by observations is of 6.6 arcmin$^2$ centered at the
Galactic coordinates $l=337^\circ.9$, $b=-11^\circ.7$. Figure
\ref{CMD} shows the $(V-I, V)$ color--magnitude diagram for the about
1000 stars whose proper motion is significantly different from that of
the NGC~6397 stellar population (cf.\ Fig.\ref{CMD} in KACP).

In Section 2 we will use these data to discuss the predictions of the
Bahcall-Soneira Galaxy model (Bahcall 1986) which has been
widely used by observers to predict the number of stars of different
types, colors and apparent magnitude ranges in different fields of
interest (see e.g. Boeshaar \& Tyson 1985, Flynn \& Freeman 1993,
Lasker et al. 1990, Stuwe 1990, Basilio et al. 1996).  Bahcall and
Soneira assumed that the Galaxy is simply composed by a disk, with a
scale height of a few hundred parsecs, and a more or less spherical
spheroid. The Bahcall-Soneira model, as further improved mainly to
account for interstellar reddening and obscuration, was proved to be
in good agreement with available observations (see e.g Bahcall 1986,
Gould, Bahcall \& Maoz 1993 and references therein) and the export
code was made available on the Web to the scientific community in
1995. However at faint magnitudes such an agreement is
strongly dependent on the assumption about the luminosity function
(LF) of faint main sequence (MS) stars. As a result, in Sect.3, we
will show that the above quoted agreement vanishes for several recent
evaluations of the actual LF of either the disk or the halo components.

The need for more complex Galactic models has been advanced in 1983 by
Gilmore \& Reid, who firstly suggested the occurrence of an extended
``thick disk'' population consisting of stars with spatial and
kinematic properties intermediate between disk and spheroid. 
Starcounts is not the best way to constrain
this third component because similar results could be obtained
either with a thick disk or without it in dependence of the adopted model
parameters (disk/thick disk scale height, scale lenght and normalizations). 
However detailed analysis which, in addition to the star counts,
take into account the  velocity distribution of
local stars, support the need for a thick disk component (Ojha et
al. 1996, Wyse \& Gilmore 1995, Norris \& Ryan 1991, Casertano,
Ratnatunga \& Bahcall 1990.)

However up to now, sensitively different values for the
thick disk structural parameters (density law, local density etc..)
have been proposed (see e.g. Reid \& Majewski 1993, Yamagata \& Yoshii
1992, Ojha et al. 1996, Mendez \& Guzman 1998).  Thus the situation is
not yet well defined and in Sect.4 we will use our data to test some
three-component models currently available in the literature. In this
context, one has to remind that with very faint observations it should
be possible to resolve the Galaxy into its component populations. This
because for V$>$19, at least for high Galactic latitudes, the various
components are expected to contribute with different colors: the bluer
colors are dominated by spheroid stars, the reddest objects do belong
to the disk population, while the intermediate colors should be
dominated by thick disk stars (see Basilio et al. 1996).

Unfortunately, our data refer to a sky region of somewhat low
latitude, which gives a sensitive gain in the number of stars but
which - as we will find - makes difficult a clear separation among the
various populations.  As a result we will find that either two-component
 or three component models give an equally satisfactory
agreement with magnitude and color observational counts.  No one of
the current Galactic models appears in agreement with observations for
some extreme assumptions on the faint LF, whereas the data do not
allow to discriminate between different three-component models within
the current uncertainties on the scale height and on the density of
stars in the solar neighbourhood.


\section{The B\&S model}

As first step we compared color and magnitude distributions from our
observational sample to the predictions of the Bahcall-Soneira model
in its original export version (see Bahcall, 1986, for a complete
description of the model). For convenience in the following we will
briefly summarize the main parameters (see also table \ref{tab1}:\\ 
i) {\em Disk}. The disk has a Wielen (1974) luminosity function and an
exponential scale lenght of 3.5 kpc. The scale height for the later
main sequence (M$_V$$>$5) is 325 pc. For the early main sequence
(M$_V$$<$5) it is (82.5 M$_V$ $-$ 97.5) pc with a minimum of 90
pc. The scale height of the giant branch is 250 pc.\\ ii) {\em
Spheroid}. The spheroid has the same LF as the disk implemented for
M$_V$$\leq$4.5 with the 47 Tuc luminosity function (Da Costa, 1982),
shifted to fit the Wielen LF at M$_V$$\approx$4.5.  The Galactocentric
distance is R$_{\circ}$=8 kpc and the axis ratio is 0.8. The density
is normalized locally at 1/500 of the disk.

For each stellar population it is possible to associate intrinsic
stellar magnitudes with colours, predicting in such a way the expected
colour distribution of stars in each given interval of magnitudes.
The original version of the B\&S model adopts as relations between
(B-V) colors and absolute visual magnitudes, a disk main sequence (MS)
from observations by Johnson (1965,1966) and Keenan (1963) and the
color-magnitude (CM) diagram of M67 by Morgan \& Eggleton (1978) for
disk giant stars. The program allows the choice of the spheroid giant
branch from a number of different color-magnitude diagrams (e.g. 47
Tuc, M13, M15, M92) while the main sequence adopted for the disk is
shifted to properly match the evolved stars of the selected diagram,
reproducing in this way the different colors of the spheroid MS as a
function of the assumed metallicity.
 
\begin{figure}
\label{mag}
\centerline{\epsfxsize= 8 cm \epsfbox{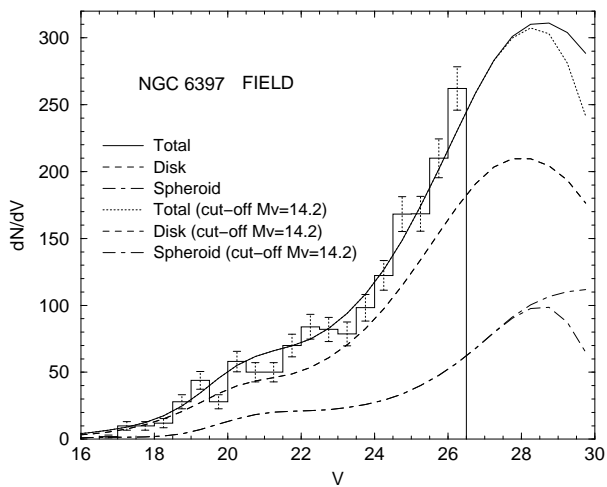}} 
\caption{The histogram of the observed magnitudes as compared with
theoretical predictions by the B\&S model.  Observations are corrected for incompleteness.
Error bars give the expected 1$\sigma$ statistical
fluctuations of the observed counts in the completeness region. Heavy
dashed and dot-dashed lines represent the contribution to the total
counts of the disk or the spherical component, respectively. Thin
dot-dashed and dotted lines show the results for the spheroid and the
corresponding total counts when the cut-off for the spheroid luminosity
function is introduced (see text).}
\end{figure}

Figure 2 compares the distribution of apparent $V\,$-magnitudes of the
sample stars with the corresponding predictions produced by a
straightforward application of the B\&S model. The observed magnitude
distribution plotted in Fig.\ref{mag} has been corrected for
incompleteness as in KACP (cf.\ Tab. 1 in KACP), restricting the
investigation to the range of magnitudes with completeness larger or
of the order of 80\%. i.e. to stars with $V\leq 26.5$ . As shown in
Fig.\ref{mag}, one finds a good agreement over the whole
accepted range of magnitudes, with theoretical predictions almost
always within the expected statistical fluctuation of the counts. As
expected, at these low latitudes, counts are dominated by disk stars.
Thus, as we will discuss better later, the total distributions
 are barely sensitive to the presence of a possible third component.

\begin{figure}
\label{col1}
\centerline{\epsfxsize= 4 cm \epsfbox{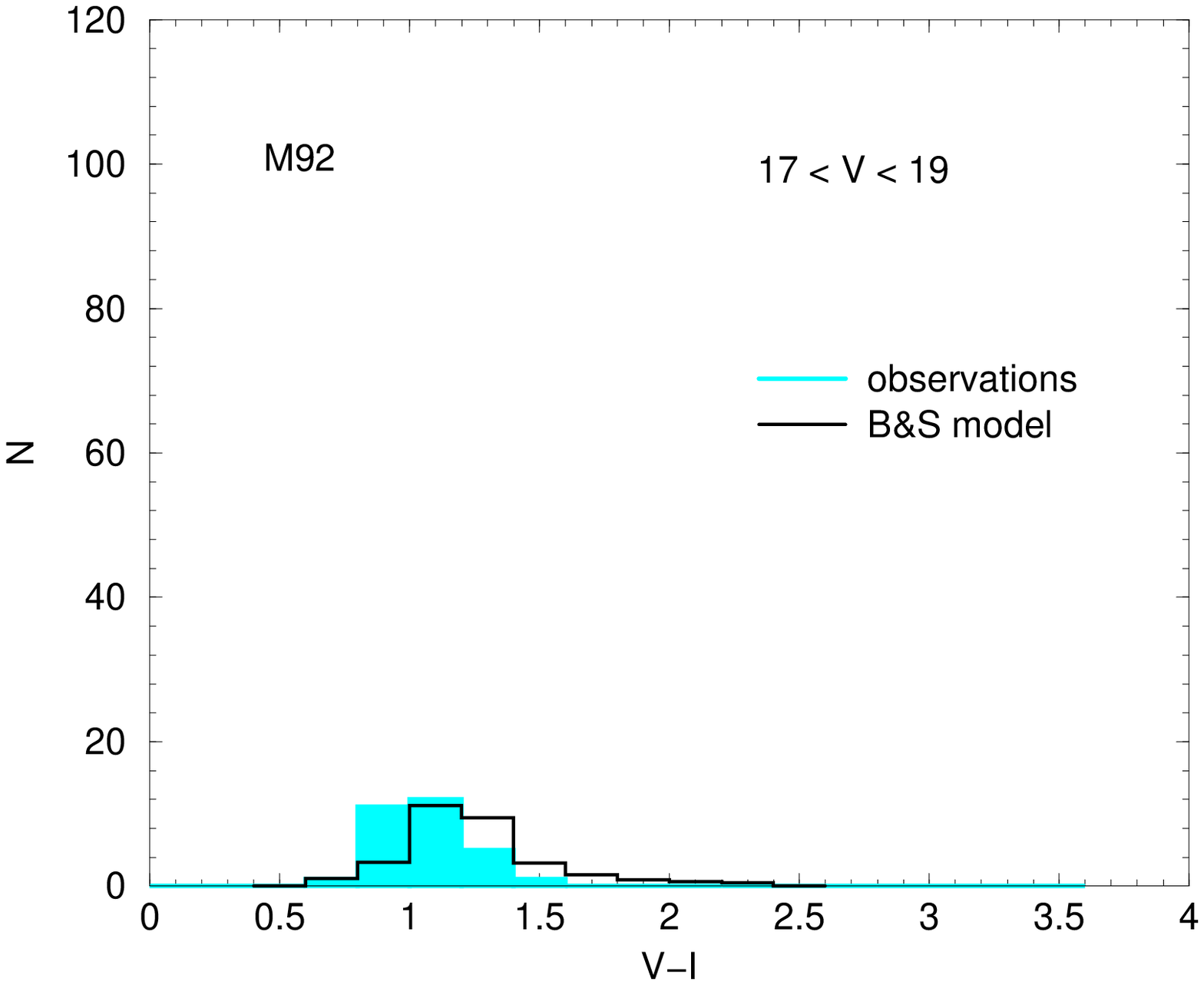}
\epsfxsize= 4 cm \epsfbox{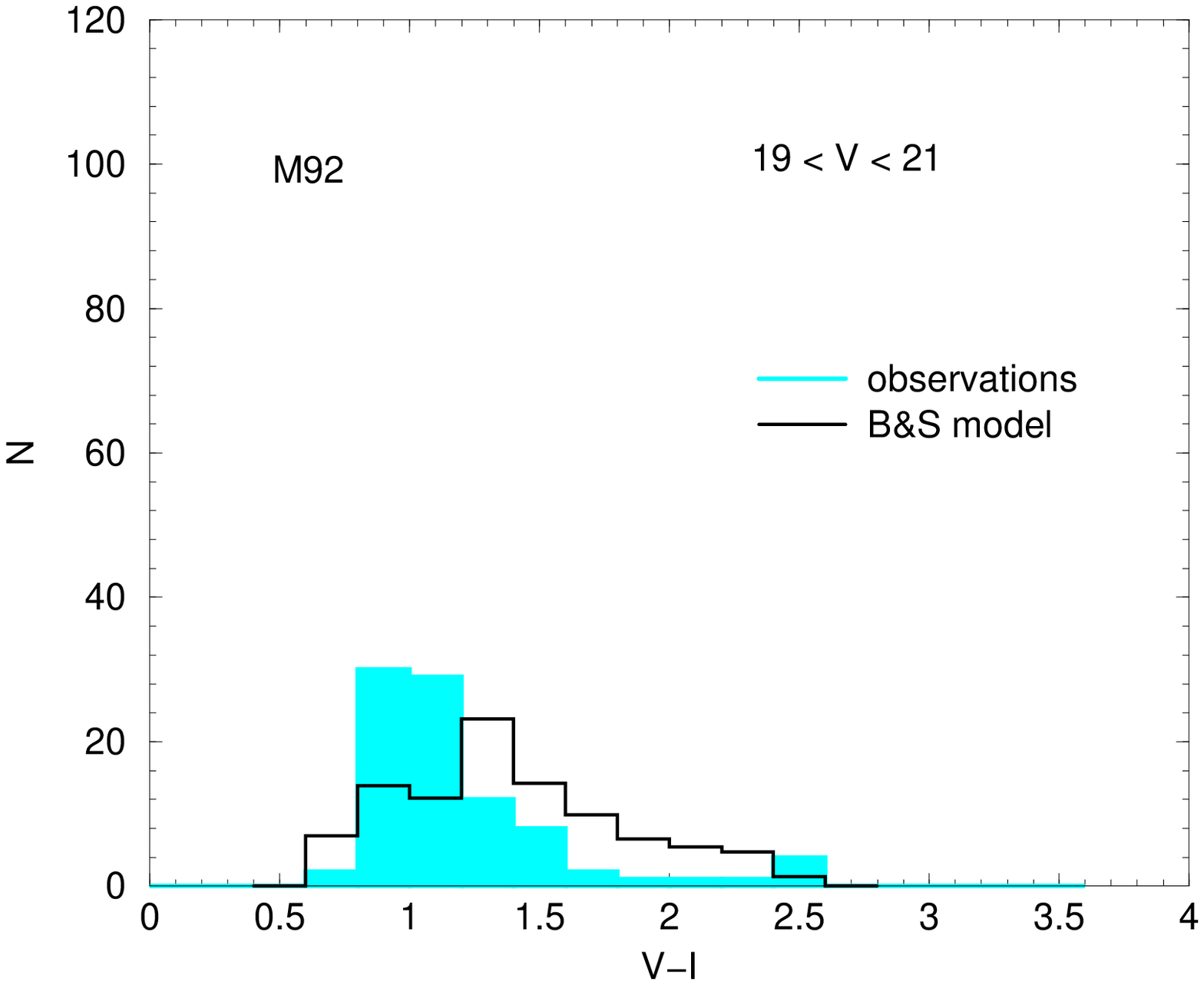}} 
\centerline{\epsfxsize= 4 cm \epsfbox{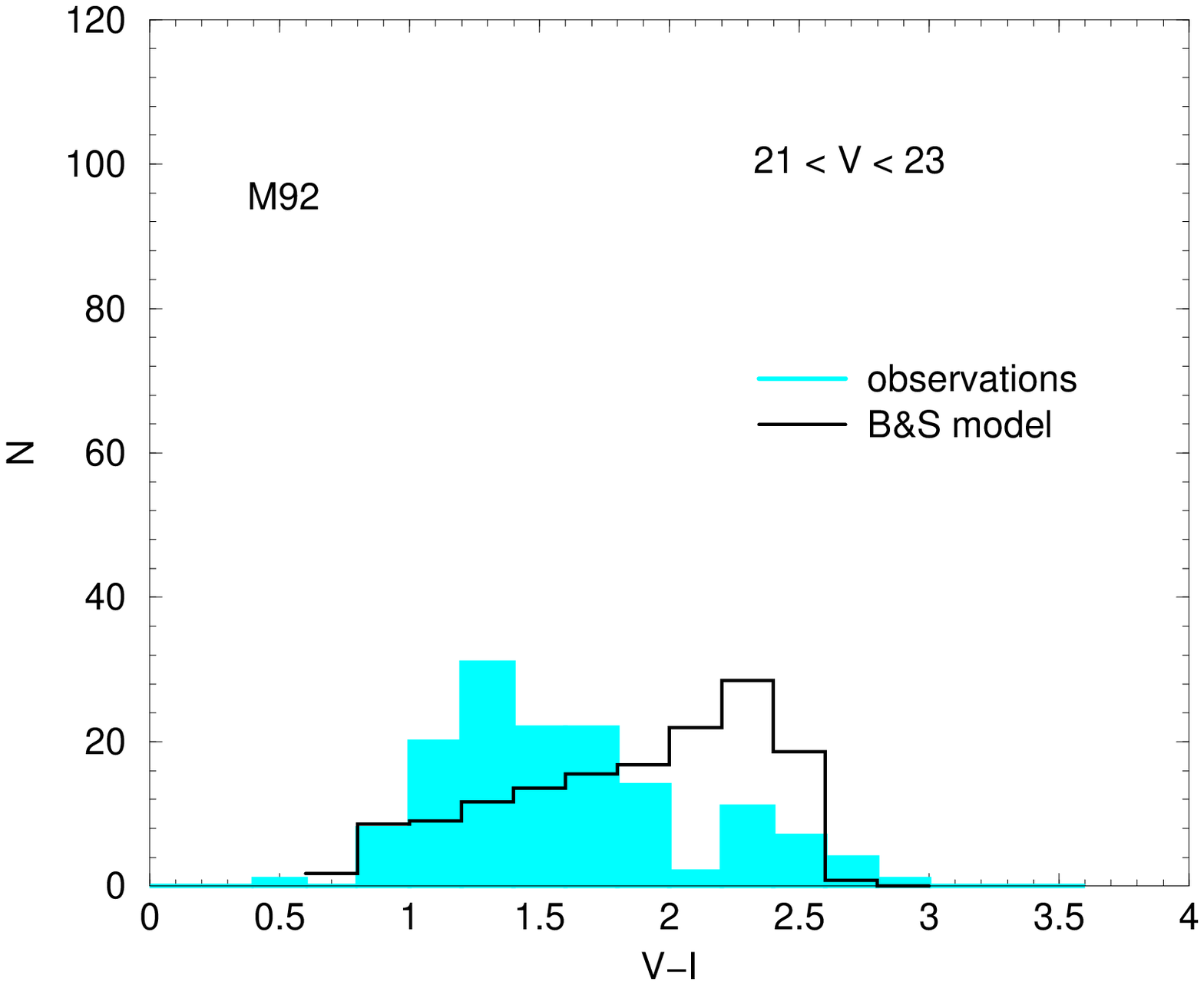}
\epsfxsize= 4 cm \epsfbox{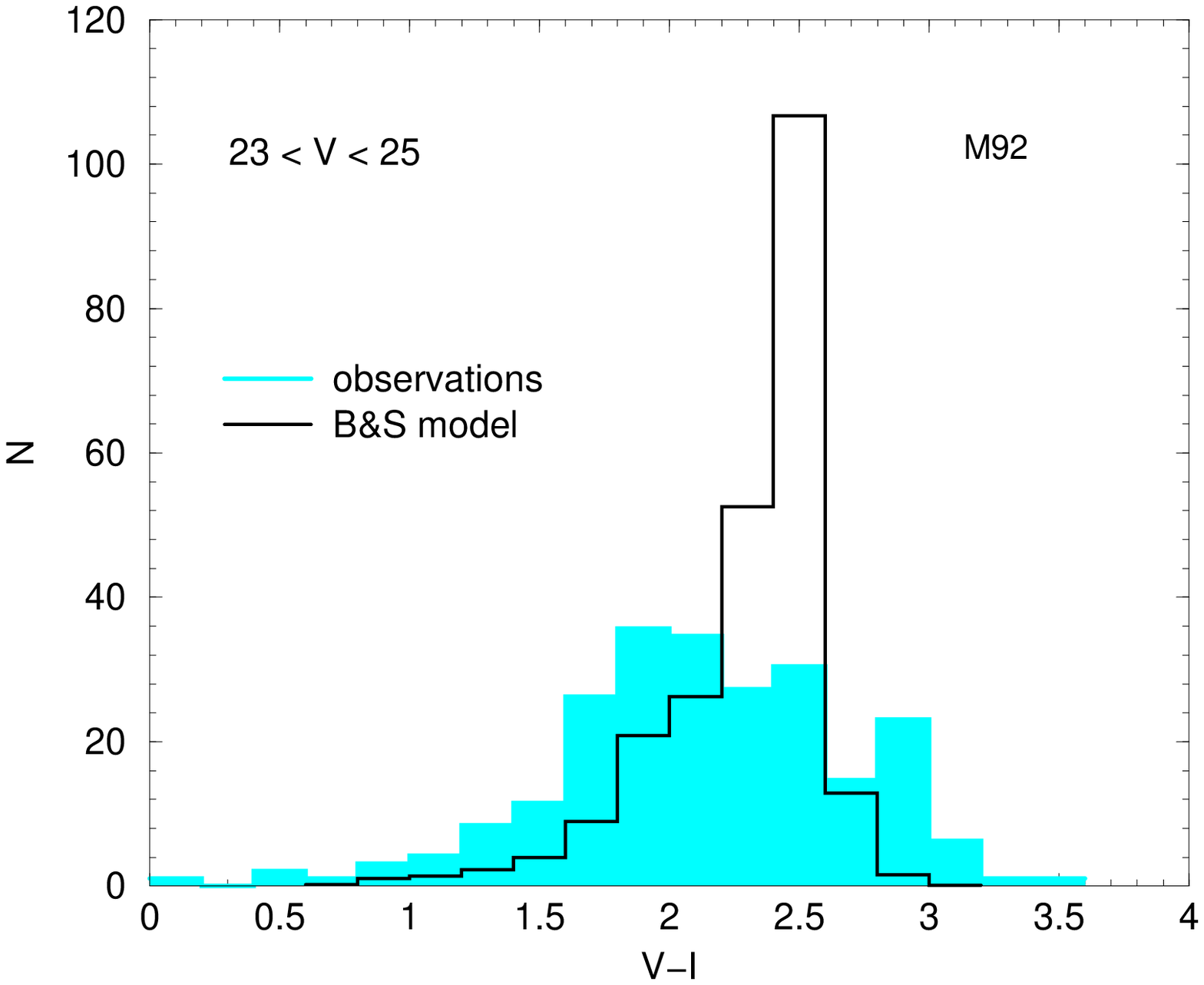}} 
\caption{Comparison between the observed (shaded region) and predicted distribution of
($V-I$) colours in the labelled intervals of magnitude. The adopted
($M_V$,$V-I$) relations for MS stars are as in Fig.4, upper panel,
while the ($M_V$,$V-I$) relations for RGB stars are derived from the
CMDs of M67 (Morgan \& Eggleton 1978) and M92 (Sandage 1970) for disk
and spheroid stars respectively. The adopted color dispersion is $\sigma$=0.06
(see e.g. Bahcall \& Soneira, 1980).}
\end{figure}

The observational data can also be used to test predicted color
distributions.  We used the M$_V$-(B-V) relations described above
together with the relations from ($B-V$) to ($V-I$) colour bands for MS
and giant stars obtained from Johnson (1965, 1966) observational data,
as suggested in the original B\&S code.
Figure 3 shows that, in this case, the agreement is far from being
satisfactory, the difference between predictions and observations
becoming worse at fainter magnitude. Thus while V magnitudes are
well reproduced, $(V-I)$ colour distributions are not. The origin of
such an occurrence is easily understood, since inspection of the B\&S
data discloses that the adopted ($M_V$, $B-V$) diagram for the two
stellar populations was still in excellent agreement with recent
results from the Hipparcos satellite for nearby stars (Kovalevsky
1998), whereas this is not the case for the ($M_V$, $V-I$)
distributions.
To be quantitative, the upper panel in Fig. 4 compares the adopted
B\&S ($M_V$,$V-I$) distributions for disk and spheroid MS stars with
recent data by Monet et al.\ (1992) and Dahn et al.\ (1995) for dwarf
and subdwarf stars in the solar neighbourhood; the need to update the
M$_V$-($V-I$) relations appears obvious.  In passing we note that the
B\&S model was never compared by the authors with ($V-I$) color
distributions thus they had not the need to introduce in the code
precise relations for this color band.  We derived a new ($M_V$,
$V-I$) relation for the disk population by best fitting the
above-cited observational data for dwarf stars.


\begin{figure}
\label{vvi}
\centerline{\epsfxsize= 8 cm \epsfbox{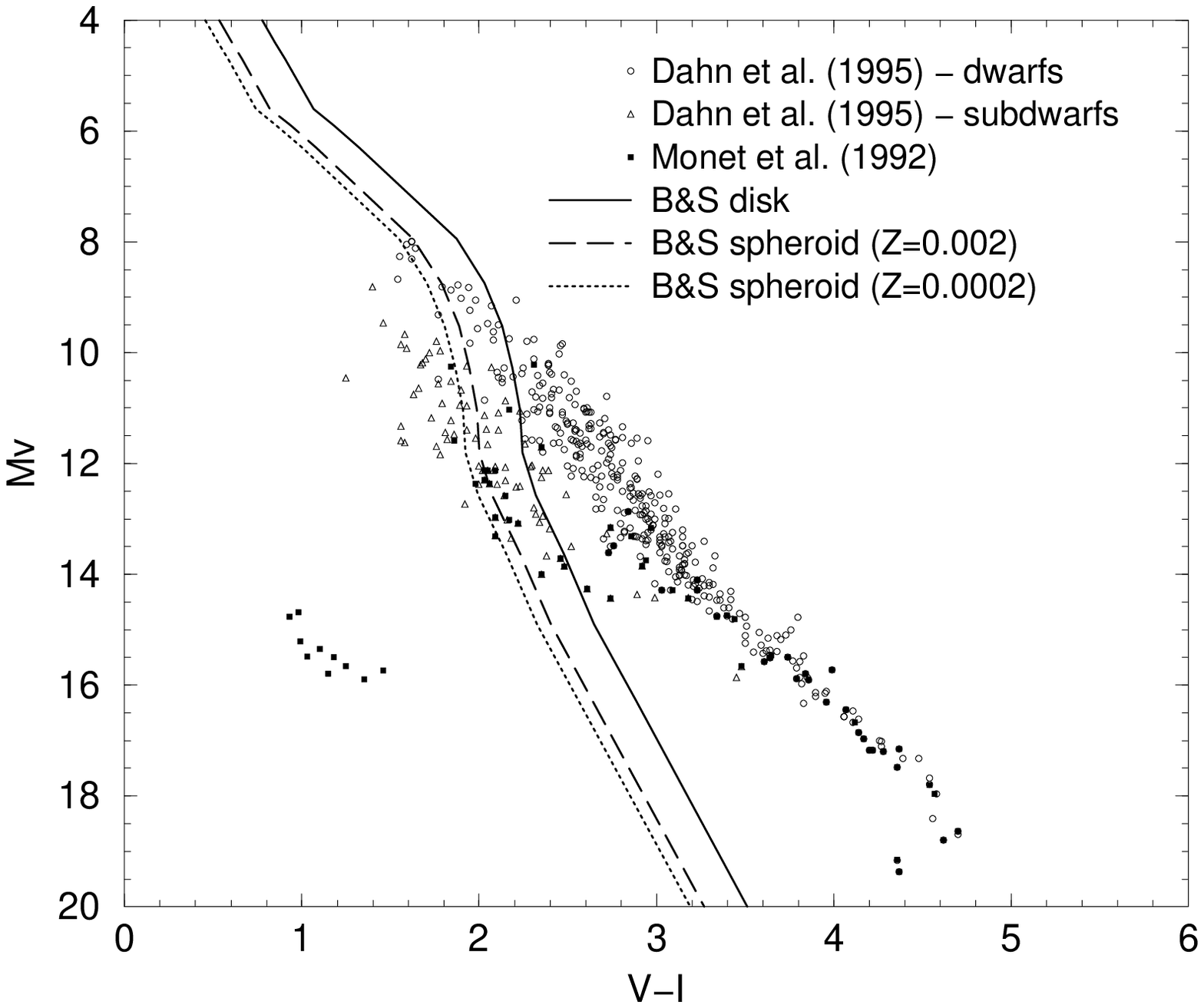}}
\centerline{\epsfxsize= 8 cm \epsfbox{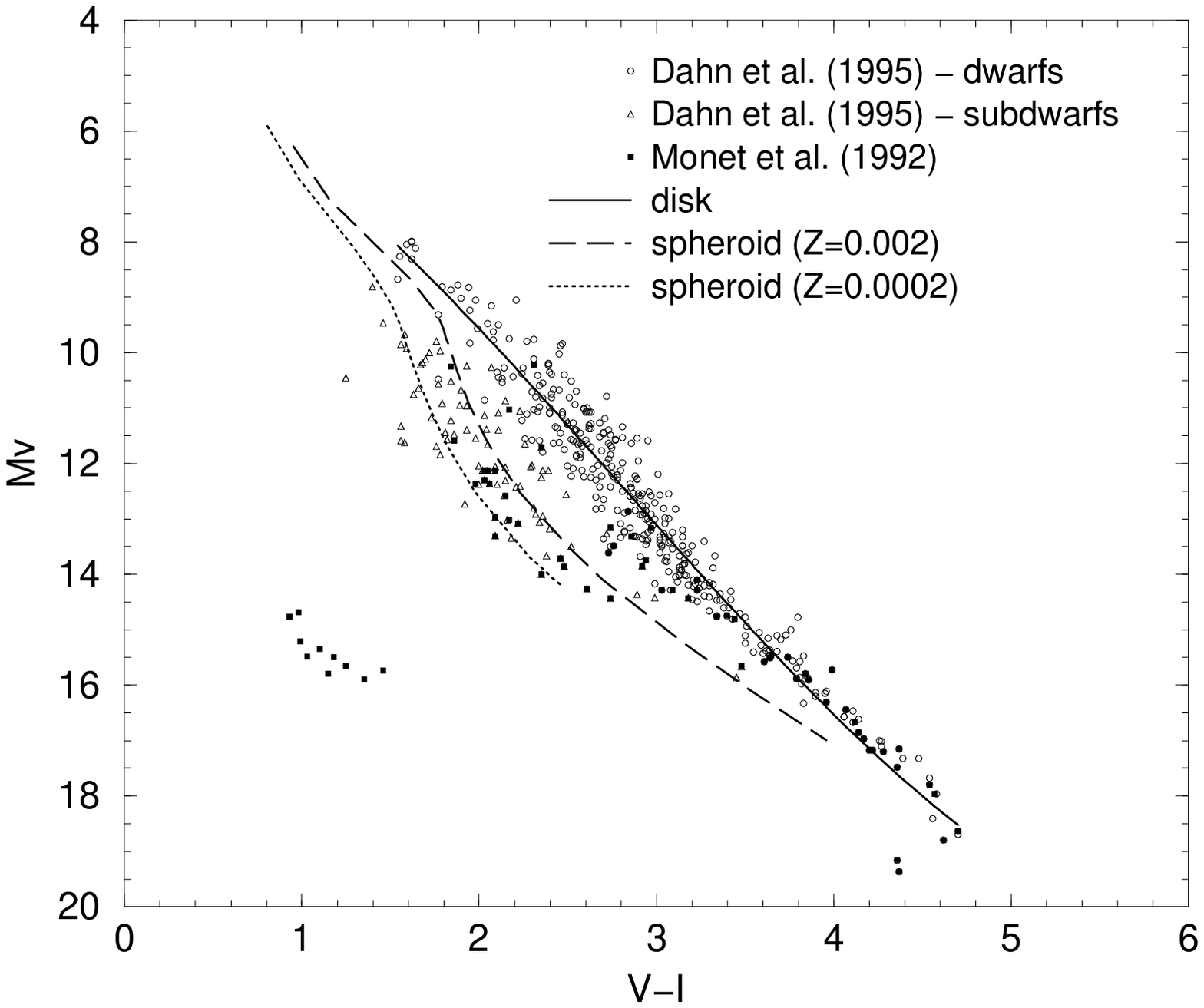}} 
\caption{Upper panel: The (M$_V$, V-I) relations adopted in
 the original B\&S model as compared with the CM diagram of stars in the
 solar neighbourhood. Lower panel: the same as in the upper panel but
 for the ``new'' (M$_V$, V-I) relations (see text).}
\end{figure}

In order to keep our computations as close as possible to the original
B\&S model, we choose for the spheroid a $Z=0.0002$ MS as defined by
HST in the very metal-poor globular M30 (Piotto et al.\ 1997). The
bottom panel of Fig.4 shows the new sequences as compared with
the same observational data in the upper panel.  Here we notice that
the excellent agreement recently found between theoretical predictions
and HST observations of faint MS in Galactic globulars (Cassisi et
al.\ 2000) supports the use of theoretical data in varying the adopted
spheroid metallicity. Thus in the lower panel of Fig.4 the low MS at Z=0.002 is from
evolutionary calculations by Cassisi et al. (2000). Finally,
($M_V$,$V-I$) relations for giants stars that have left the MS have
been taken from the CM diagrams of M67 (Montgomery, Marshall \& Janes
1993) and M92 (Johnson \& Bolte 1998) for disk and spheroid stars,
respectively.

However, even if relying on observational data, one can
not ignore that all the available theories unanimously predict that
the location of the lower mass limit for H burning is strongly
dependent on the star's metallicity and, in particular, that for
$Z=0.0002$ this limiting magnitude is $M_V=14.2$ (see, e.g., again
Cassisi et al.\ 2000). To be realistic, we thus used this magnitude as
a cut-off of the luminosity function of the spheroid
component. Numerical experiments, as reported in the previous Fig.2,
shows that such a cut-off is predicted to affect the distribution of
the observed luminosities only at magnitudes fainter than $V=28$, thus
beyond the limit of the present investigation.  However, it
appears interesting to notice that, at least in principle,
observational data below $V\sim$28 should constrain the magnitude of
the cut-off, and, therefore, the mean metallicity of the spheroid
component.

Before comparing theory and observation with the new color-magnitude
relations, it is useful to discuss the adopted reddening correction.
The original version of the code incorporates for intermediate-to-high
Galactic latitudes ($\mid b\mid\geq$10$^{\circ}$) the ``default''
reddening/obscuration correction through ``infinity'' from Sandage
(1972) based on Galaxy counts, which gives in the direction of NGC 6397
a reddening E(B-V)$\approx$0.26. However we prefer to adopt the more
precise and extensively used reddening maps by Burnstein \& Heiles
(1982).  For their reddening estimates the authors used together the
results of surveys of neutral hydrogen column densities and of deep
Galactic counts. The Burnstein \& Heiles compilation gives E(B-V)$\approx$0.18
in the direction of NGC 6397, a value  in good agreement with
reddening estimates for the cluster from different methods:
E(B-V)=0.18 $\pm$0.02 (see e.g. Harris 1996, Anthony-Twarog et
al. 1992, Drukier et al. 1993).  Thus, in the following, we will adopt
E(B-V)=0.18 transformed in E(V-I) by using E(V-I)$\approx$1.25 E(B-V)
by Bessel \& Brett (1988). The reddening scale height adopted in the
program is 100 pc (see e.g. Mendez \& van Altena, 1998). 

The difference between Sandage (1972) results and these estimates for the
reddening/obscuration correction clearly affects both magnitude and
colors distributions in a small but not negligible way; the magnitude
distribution is shifted by about 0.2 mag. toward brighter magnitudes
due to the reduced extinction, while the color distribution in the
interval 23$<$V$<$25 is shifted toward bluer colors by about 0.2 mag.
and the total counts in this interval are increased due to the change
in the extinction.  In passing we note that the NGC~6397 field is just
above the lower limit ($\mid b\mid=10^\circ$) of Galactic latitudes
covered by the reddening treatment by Burnstein \& Heiles (1982).

Figure 5 shows the effect of improving both the reddening and ($M_V$,
$V-I$) relations for the predicted colour distributions in the same
magnitude interval as in Fig.3. The agreement now appears rather
satisfactory, showing that with such a minor modifications the B\&S model
appears able to fit all the observational constraints reasonably
well. Numerical simulations show that results do not change
significantly if we assumed Z=0.002 as spheroid mean metallicity.


\begin{figure}
\label{col2}
\centerline{\epsfxsize= 4 cm \epsfbox{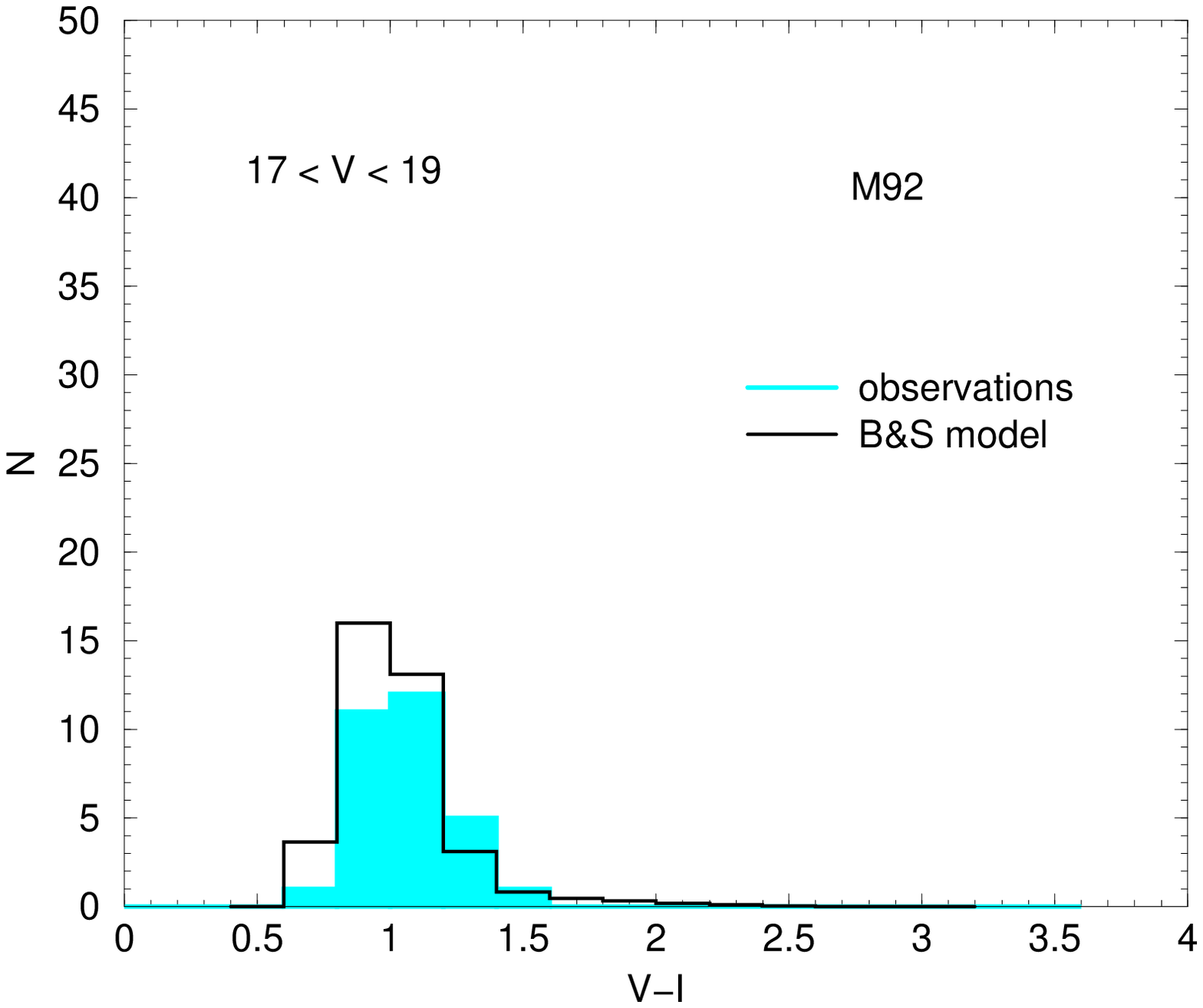}
\epsfxsize= 4 cm \epsfbox{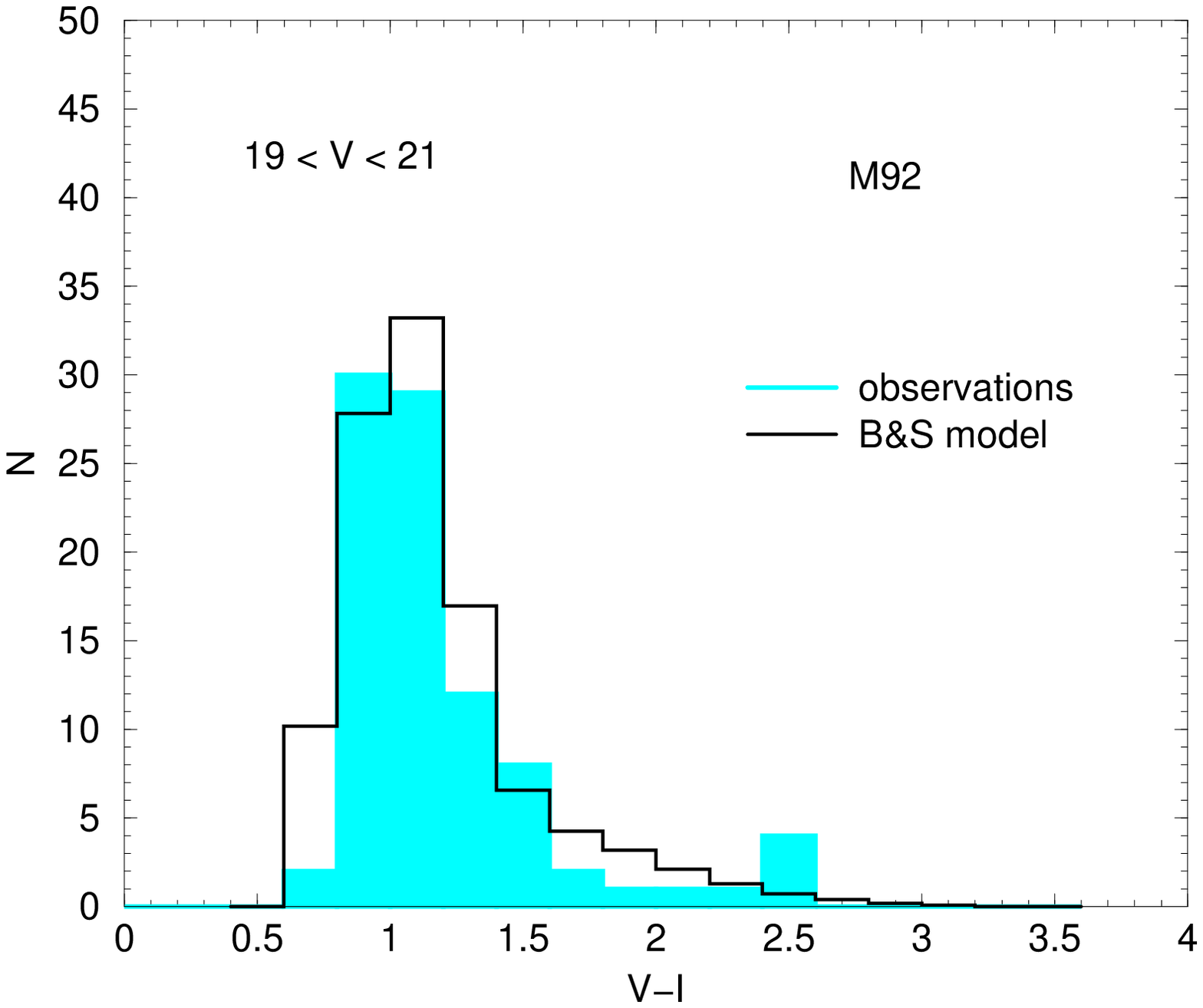}} 
\centerline{\epsfxsize= 4 cm \epsfbox{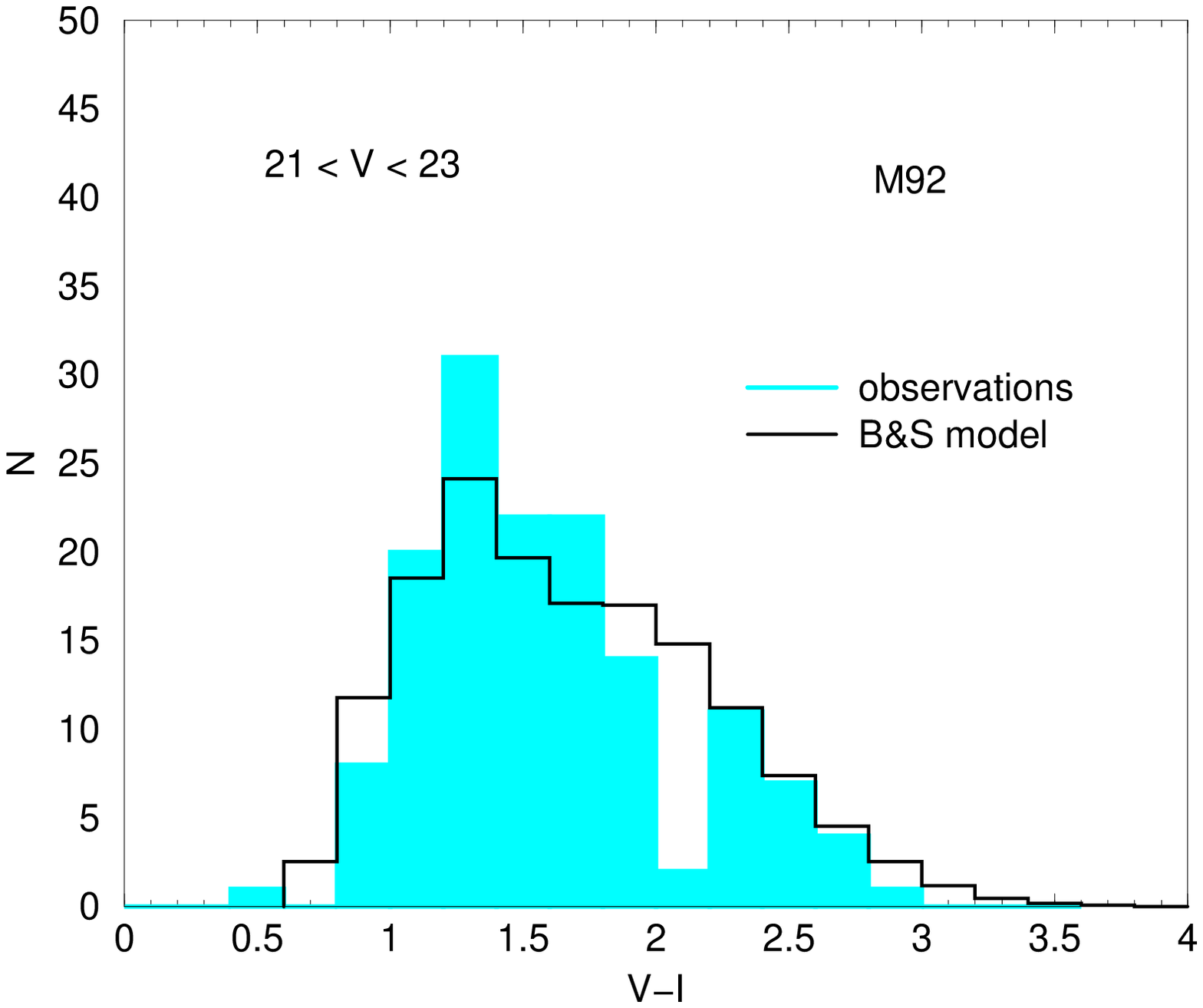}
\epsfxsize= 4 cm \epsfbox{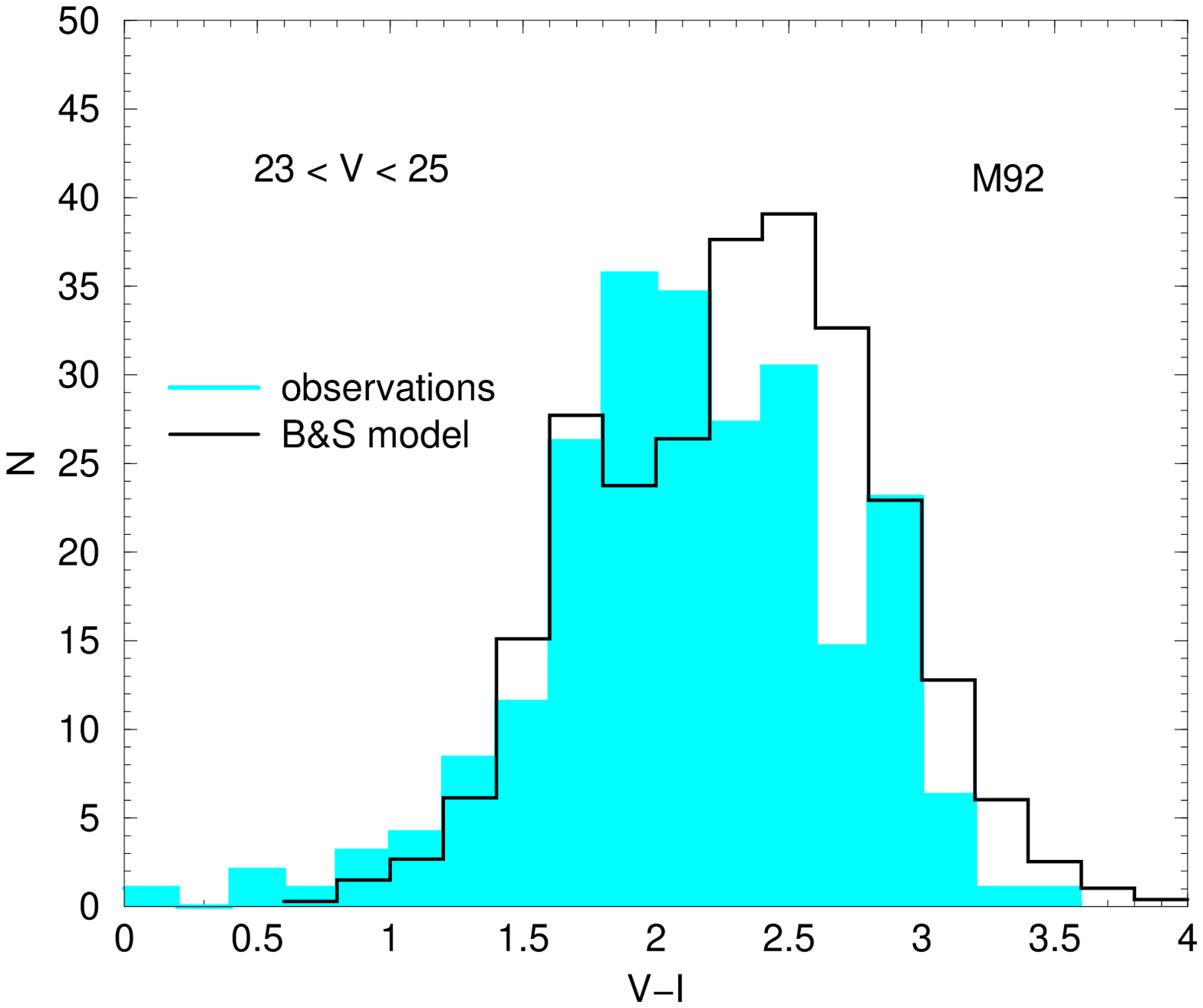}} 
\caption{Comparison between the observed and predicted $V-I$
color distributions for the labelled intervals of magnitude. The
adopted ($M_V$,$V-I$) relations for MS stars are as in Fig.4, lower
panel, while for the ($M_V$,$V-I$) relation for spheroid RGB stars we
adopted the M92 color-magnitude diagram from Johnson \& Bolte (1998).
The adopted reddening is E(V-I)=1.25 E(B-V), with E(B-V)=0.18, see text.}
\end{figure}

\section{The faint star problem}

The excellent agreement between theory and observation discussed in
the previous section is not without problems.  At present time one has
much more information about the faint end of the disk luminosity
function than at the time when the original B\&S model was
produced. As a matter of fact, below $M_V$=12.5 the original version of
the B\&S model assumes a flat luminosity function both for the disk
and the spheroid component.  However, several lines of evidence have
been found indicating that in both populations the luminosity
functions undergo a noticeable decrease at the faintest luminosities.
Figure 6 (upper panel) shows recent observational data for the faint end
of the disk luminosity function together with the distribution originally
assumed in the B\&S model.


\begin{figure}
\label{LFs}
\centerline{\epsfxsize= 8 cm \epsfbox{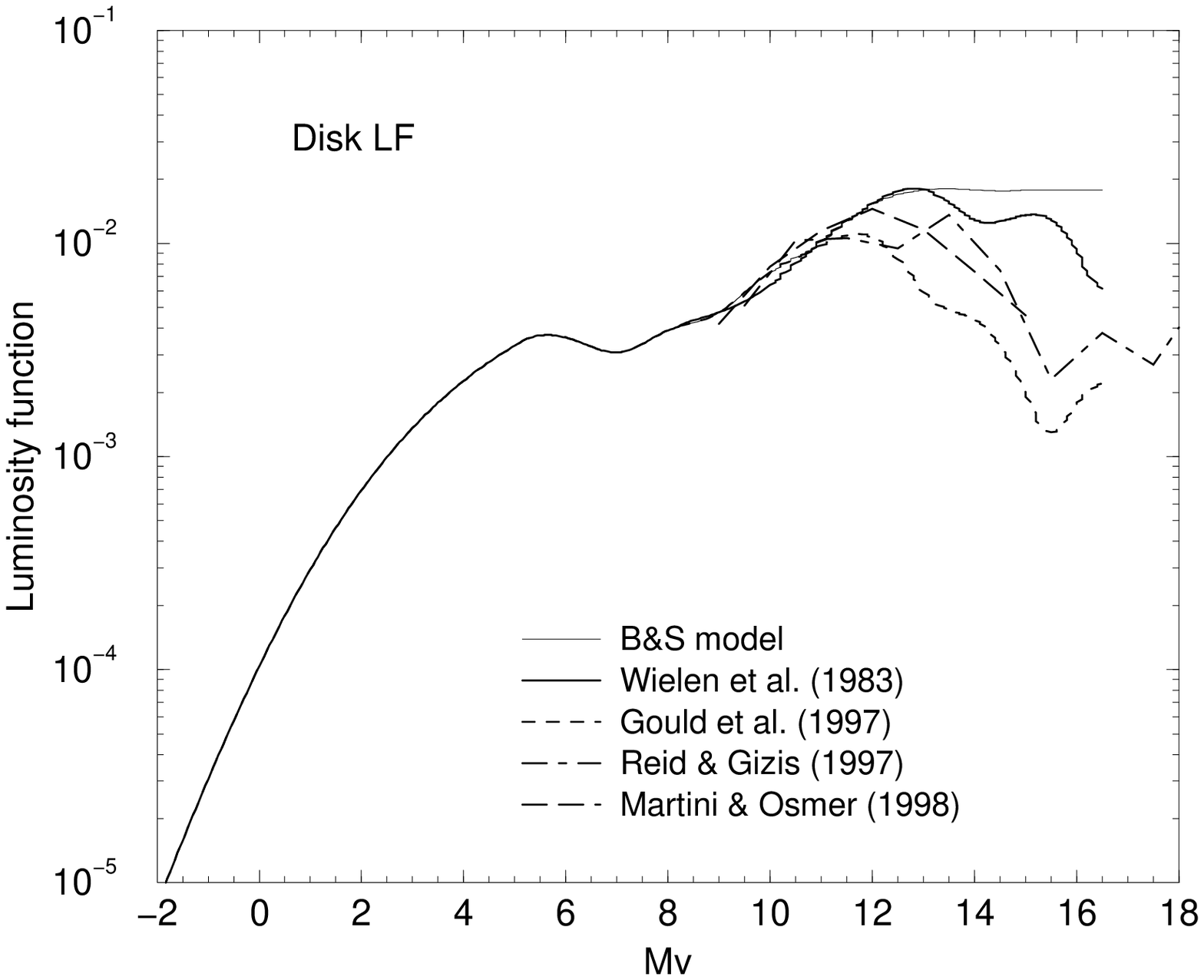}}
\centerline{\epsfxsize= 8 cm \epsfbox{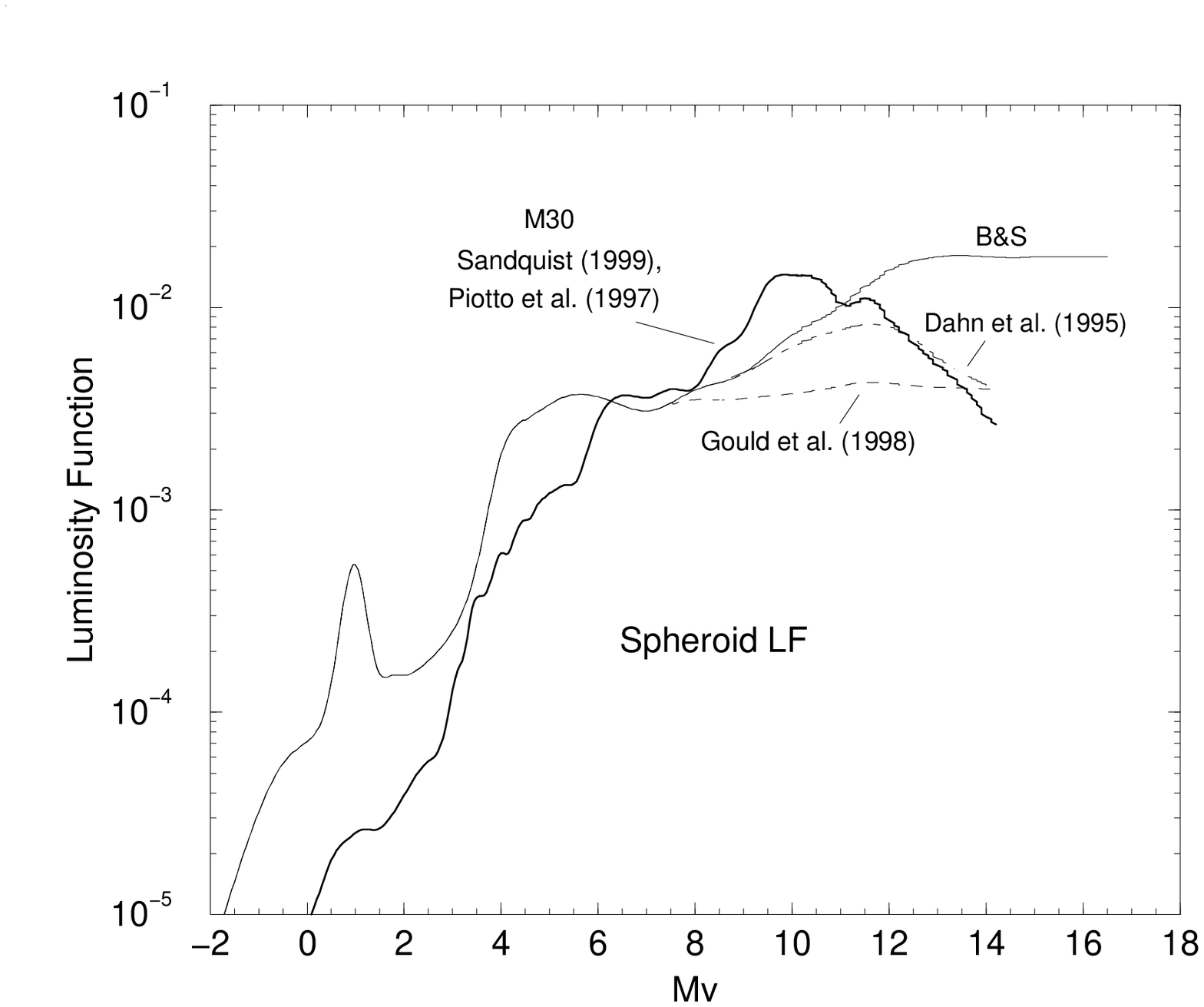} }
\caption{Upper Panel: Disk luminosity functions as determined by Wielen
et al. (1983) heavy solid line, Gould et al. 1997 (heavy short dashed line), Reid
\& Gizis 1997 (dot-dashed line) and Martini \& Osmer 1998 (long dashed
line). The original B\&S choice for the disk LF (Wielen 1974 and a
constant value for M$_V$$<$12.5) is also shown (thin solid line). Lower Panel: Spheroid
luminosity functions as determined by Gould et al. 1998 (dashed line)
and Dahn et al. 1995 (dot-dashed line) together with M30 LF from
Sandquist (1999) and Piotto et al. (1997), observational data (heavy
solid line). The original B\&S choice for the spheroid LF (the same
as the disk LF for M$_V$$\geq$4.5 implemented with the 47 Tuc luminosity
function as measured by Da Costa (1982) for M$_V$$\leq$4.5) is also shown
(solid line).}
\end{figure}

A decrease of the luminosity function beyond $M_V$= 13 has been
firstly suggested by Wielen, Jahreiss and Kruger (1983, WJK), on the
basis of parallax-star studies. An even more drastic decrease,
starting at $M_V$= 12, was more recently suggested by Gould, Bahcall
and Flynn (1996, 1997) from an analysis of HST data. The results from a
large-area multicolor survey by Martini and Osmer (1998) and those
from photometric parallax surveys by Reid \& Gizis (1997) appear
intermediate between the Wielen et al. and the Gould et
al. results. All these luminosity functions fit the Wielen et al. 
LF at a visual magnitude of about 9.

Figure 6, lower panel, shows the situation for the spheroid
component. Again there is not agreement among the results of different
authors.  Dahn et al.\ (1995) from parallaxes studies of nearby subdwarf stars
found a peak of the luminosity function at $M_V$ $\approx$
11.5. However, Gould, Flynn and Bahcall (1998) investigated HST data
and suggested a flat LF below $M_V$= 7, while very recently Gizis \&
Reid (1999), by using a sample of stars from the Palomar Sky Surveys,
support the luminosity function results by Dahn et al.\ (1995).  The
original B\&S choice for the spheroid LF is also shown.
 
An extensive investigation of the luminosity functions in Galactic
globulars has recently shown the common occurrence of a maximum near
$M_V\sim$10 (Piotto, Cool \& King 1997, Piotto \& Zoccali 1999). In
the same panel of Fig. 6 we show the luminosity function of the metal-poor
Galactic globular M30, as obtained down to $M_V$$\sim$8 by Sandquist
et al.\ (1999) and extended below $M_V$$\sim$8 with the LF of M30
provided by Piotto et al.\ (1997).  M30 is a metal poor cluster that
we will assume  as possibly representative of the spheroid
luminosity function.  The rationale for this choice follows from the
assumption of a substantial similarity between field and cluster
spheroid populations. In this context, one may take note of the
different LFs observed in Galactic globulars (see, e.g., Piotto \&
Zoccali 1999) as an evidence for a different efficiency of the
evaporation of low-mass stars from the clusters, a mechanism which is
obviously impoverishing the faint end of the original LFs. By adopting
M30 we choose one of the globulars with the largest abundance of faint
stars, thus with a LF which should be closer to the original one for
spheroid stellar content.
 
Dahn et al.(1995) and Gould et al. (1998) LFs rejoin each other
at M$_V$$\approx$ 7.5 mag. to match the
B\&S LF. M30 LF has been normalized to fit at M$_V$$\approx$7 the
spheroid LFs (see Gould et al. 1998). In passing we note that the
difference between the M30 luminosity function and the original choice
of B\&S for M$_V$$<$6 is mainly due to the normalization chosen by the
authors: the luminosity function of 47 Tuc (for M$_V$$<$4.5) has been
shifted to connect to the Wielen LF at M$_V$=4.5.
However numerical experiments show that, due to the poor population of
the high luminosity LF regions, the quoted differences for M$_V$$<$6
affects only slightly magnitude and color distributions, while
difference in the LF for M$_V$$<$4.5 does not influence at all the
results. In the following we will choose to adopt the M30 LF at brigh
magnitudes (M$_V$$<$6).

We are now in the position to explore the role played on the
predicted star counts by the faint portion of the luminosity
function.  Fig.7 shows the predicted magnitude distribution of
disk and spheroid stars when different LFs have been adopted. For
comparison the histogram of the observed counts in the field of NGC
6397 is also shown. These results are easily understood on the basis
of the shape of the adopted LFs. 


\begin{figure}
\label{diskspher}
\centerline{\epsfxsize= 8 cm \epsfbox{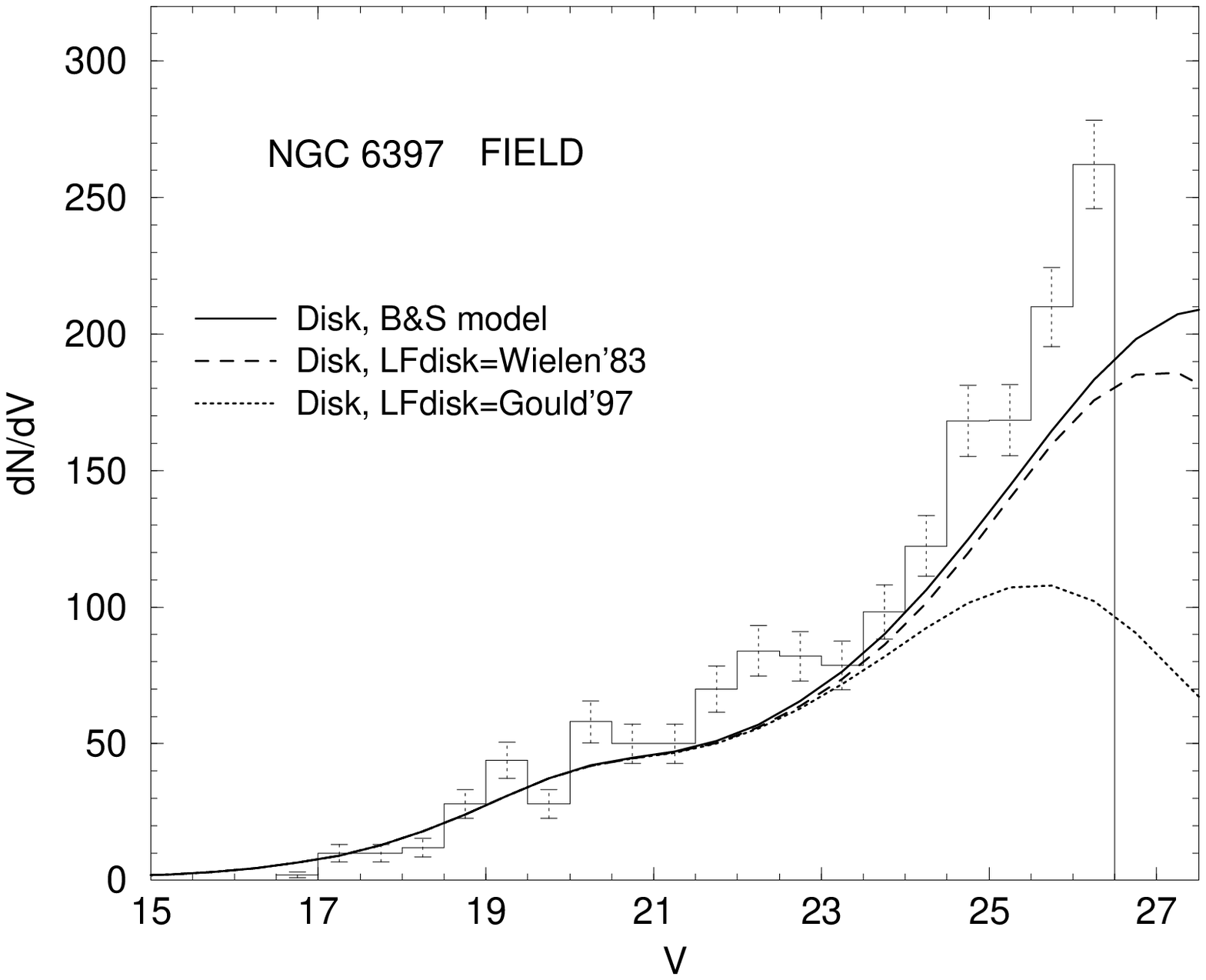}}
\centerline{\epsfxsize= 8 cm \epsfbox{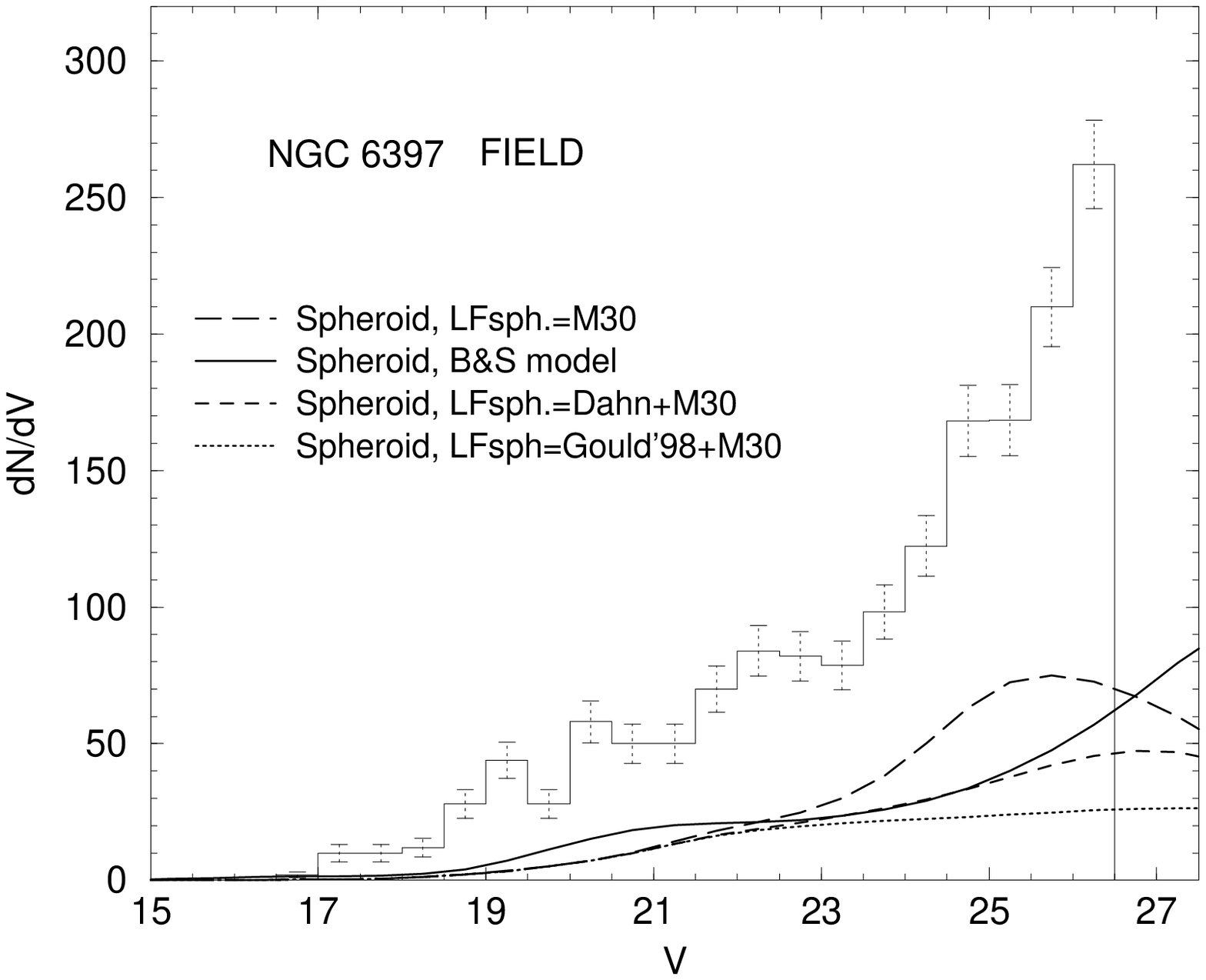}}
\caption{Upper Panel: Theoretical magnitude distribution for disk stars
when different disk LFs of Fig.6 (upper panel) are used.  Lower Panel:
Theoretical magnitude distribution for spheroid stars when the
spheroid LFs of Fig.6 (lower panel) are used.  The adopted reddening
is E(B-V)=0.18. The spheroid/disk local normalization is 1/500.}
\end{figure}

From the results of Fig.7 one also easily understand the
magnitude distribution for the different combinations of disk
and spheroid LFs shown in Fig.8.  As a first result, one finds that
different assumptions about the faint end of the luminosity function
affect star counts only for V $\gcu$ 23 (see Fig.7 and Fig.6). If one
takes the already known capability of the B\&S model to pass the
observational tests down to $V$$\sim$ 21  as an evidence that
the model gives a satisfactory description of the Galactic
distribution of the most luminous components of the stellar
populations, one would conclude that the abundance of fainter stars
depends only on the overall behaviour of the LFs.

By assuming the Gould et al. (1996, 1997) disk LF together with the Gould
et al. (1998) spheroid LF (thin solid line in Fig. 8) there is no way
to reconcile the predicted and observed $V$-magnitude
distributions. Since star counts are governed mainly by the disk
component, by assuming the Gould et al. (1997) disk LF one predicts a
number of faint stars smaller than observations in the last two
magnitude bins, even assuming for the spheroid component the M30 LF,
which gives the largest counts at faint magnitudes (long dashed line
in Fig.8).  Similarly by adopting the Gould et al.  (1998) spheroid LF
the model underestimates the observed counts in the last two bins of
magnitude for any choice of the disk LF (dot-dashed line and thin
solid line in Fig.8).

 However  the
reliability of star counts in the last magnitude bin:
26$<$M$_V$$<$26.5 is important to evaluate if the Gould et
al. (1997) LF for the disk or the Gould et al. (1998) LF for the halo are
separately acceptable for our field. We note that in the last bin (from M$_V$=26 to
M$_V$=26.5) the completeness falls from about 90\% to about 80\%.  If,
very conservatively, we decide do not take into account the results
in this bin of magnitude we cannot exclude the separate adoption
either of the Gould et al. (1997) LF for the disk or of the Gould et al. (1998) LF
for the halo, excluding however the contemporary adoption of both LFs.


\begin{figure}
\label{mvLF}
\centerline{\epsfxsize= 8 cm \epsfbox{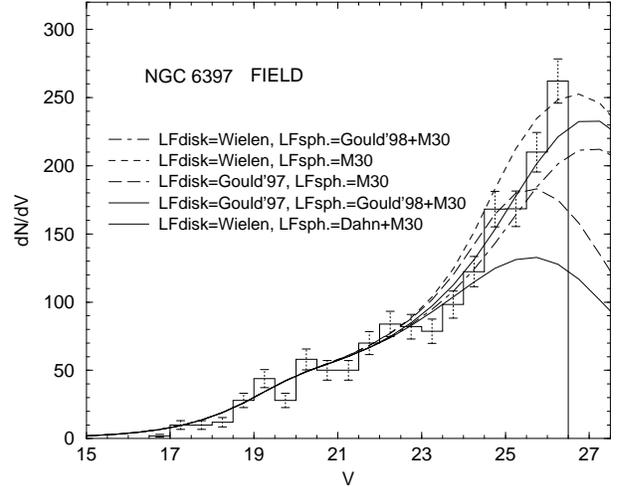}}
\caption{As in Fig.2 but for the theoretical magnitude distributions
obtained for the two-component B\&S model with the labelled
combinations of disk/spheroid LFs. 
}
\end{figure}


\begin{figure}
\label{colLF}
\centerline{\epsfxsize= 4 cm \epsfbox{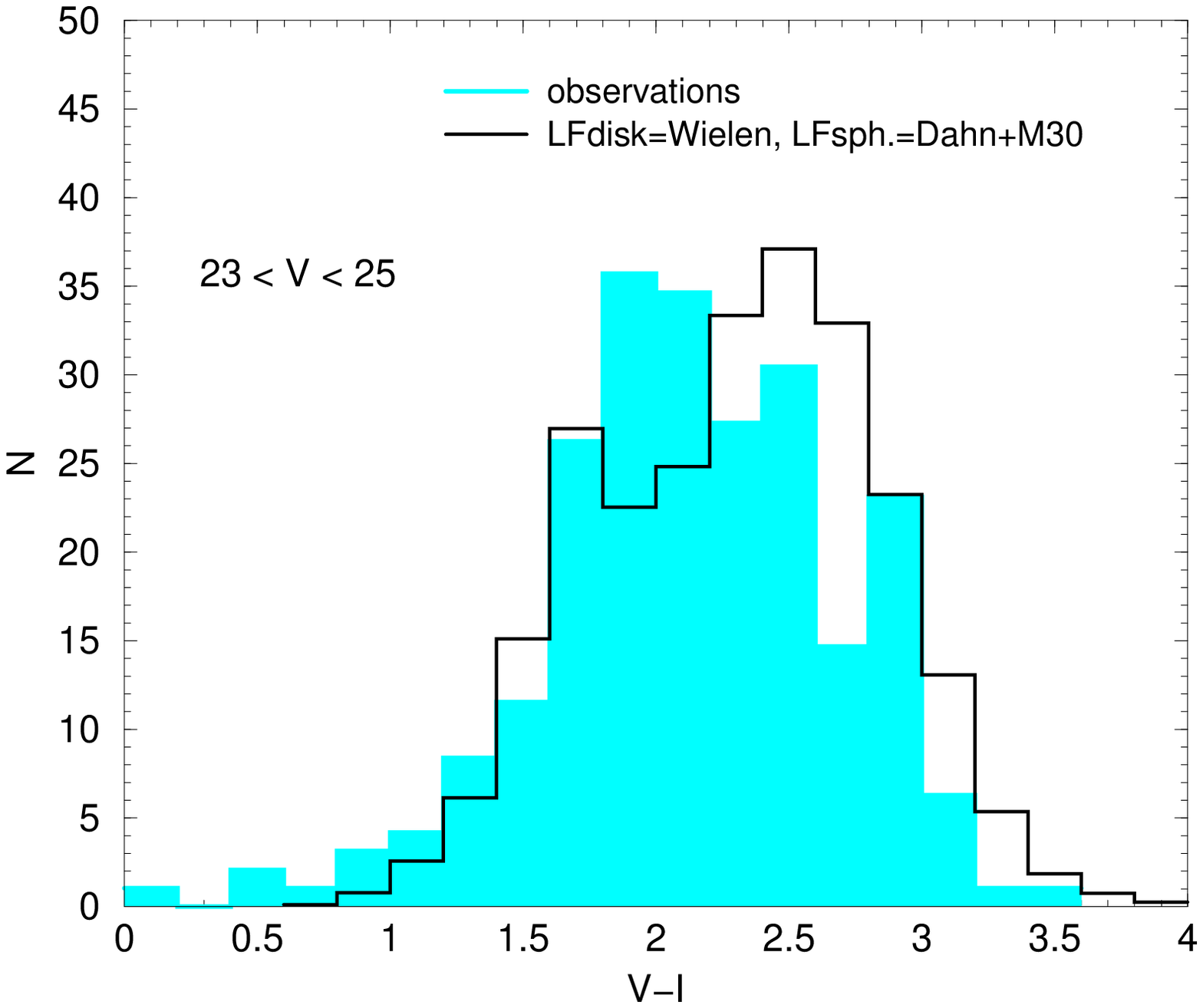}
\epsfxsize= 4 cm \epsfbox{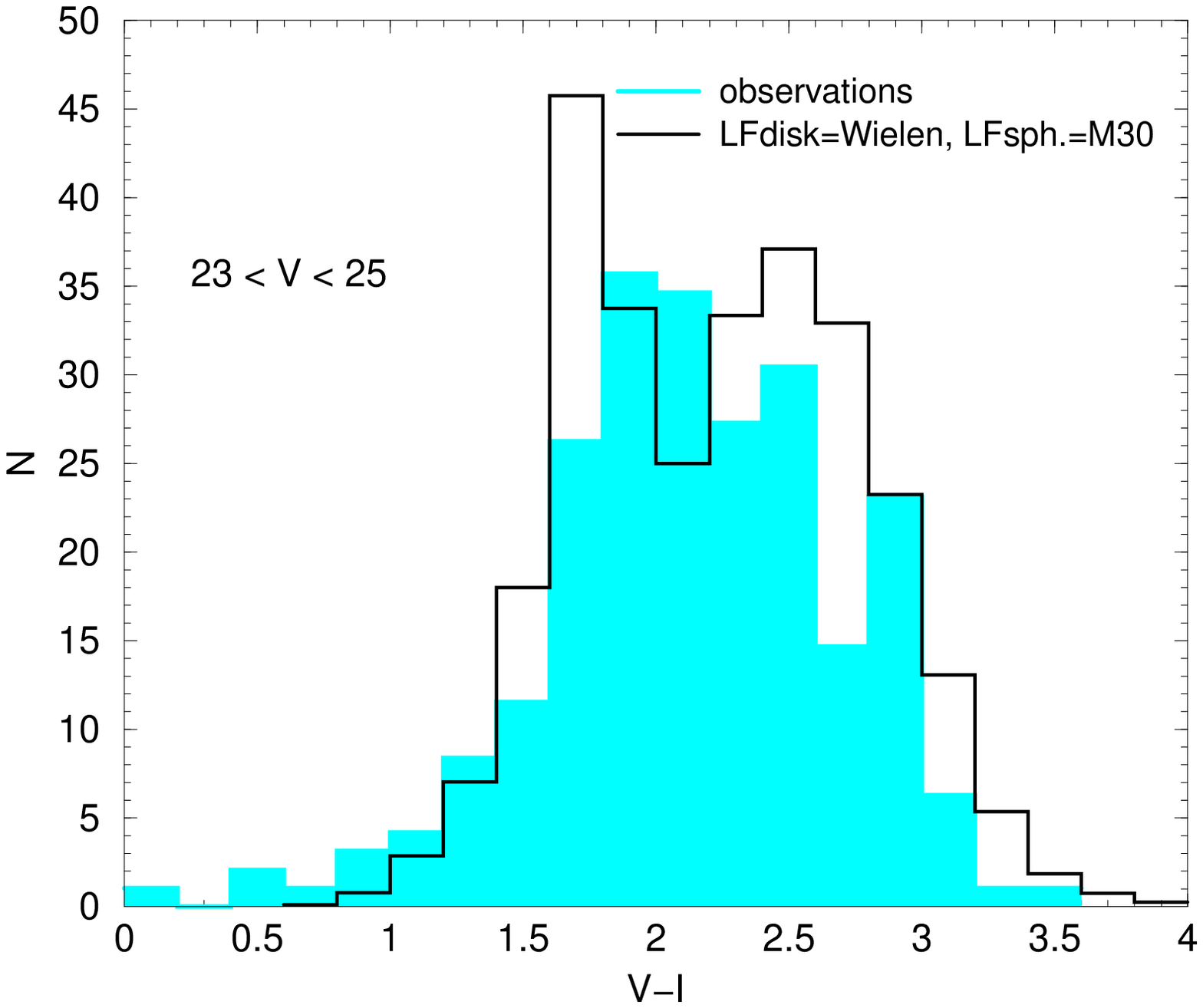}}
\centerline{\epsfxsize= 4 cm \epsfbox{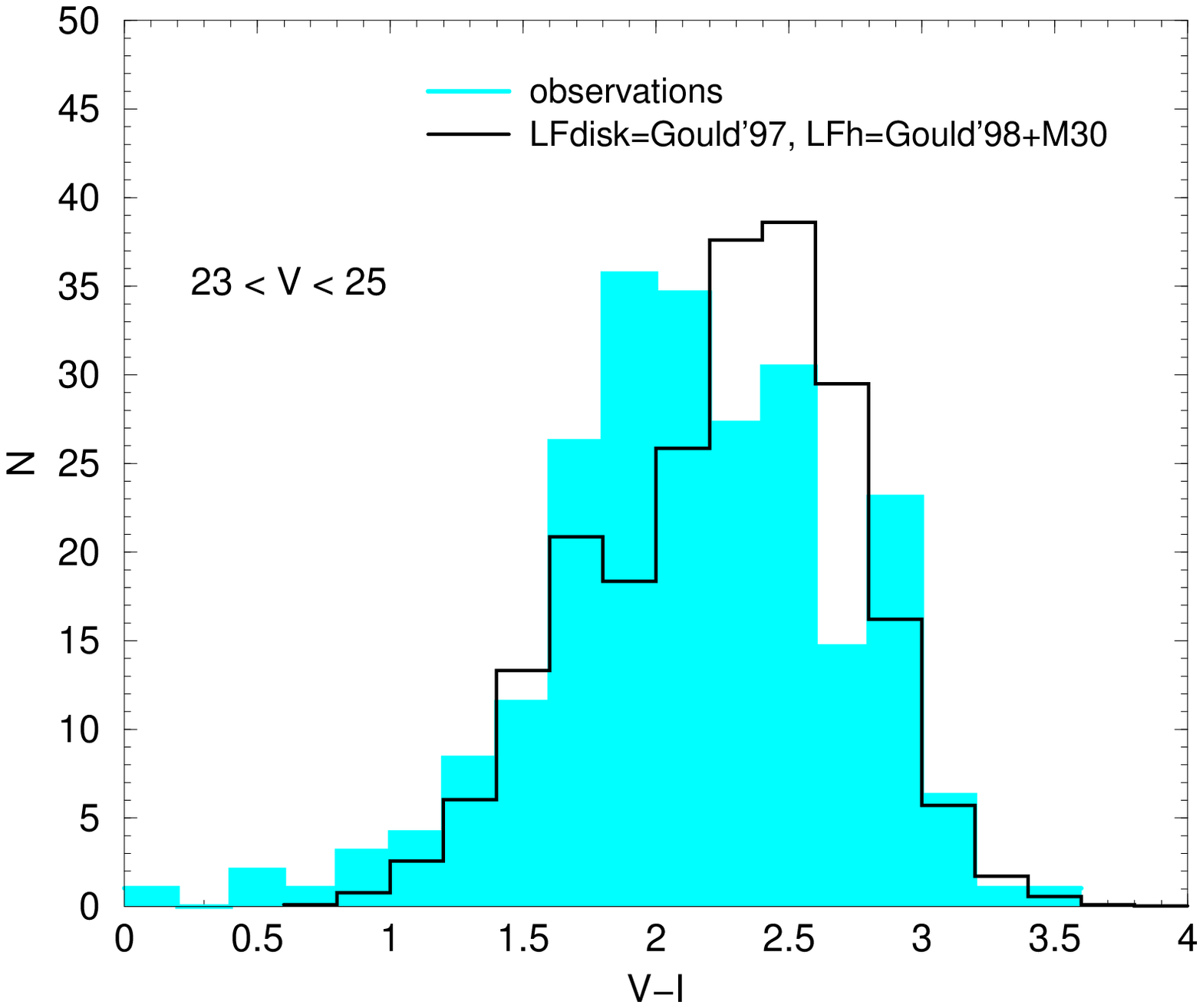}
\epsfxsize= 4 cm \epsfbox{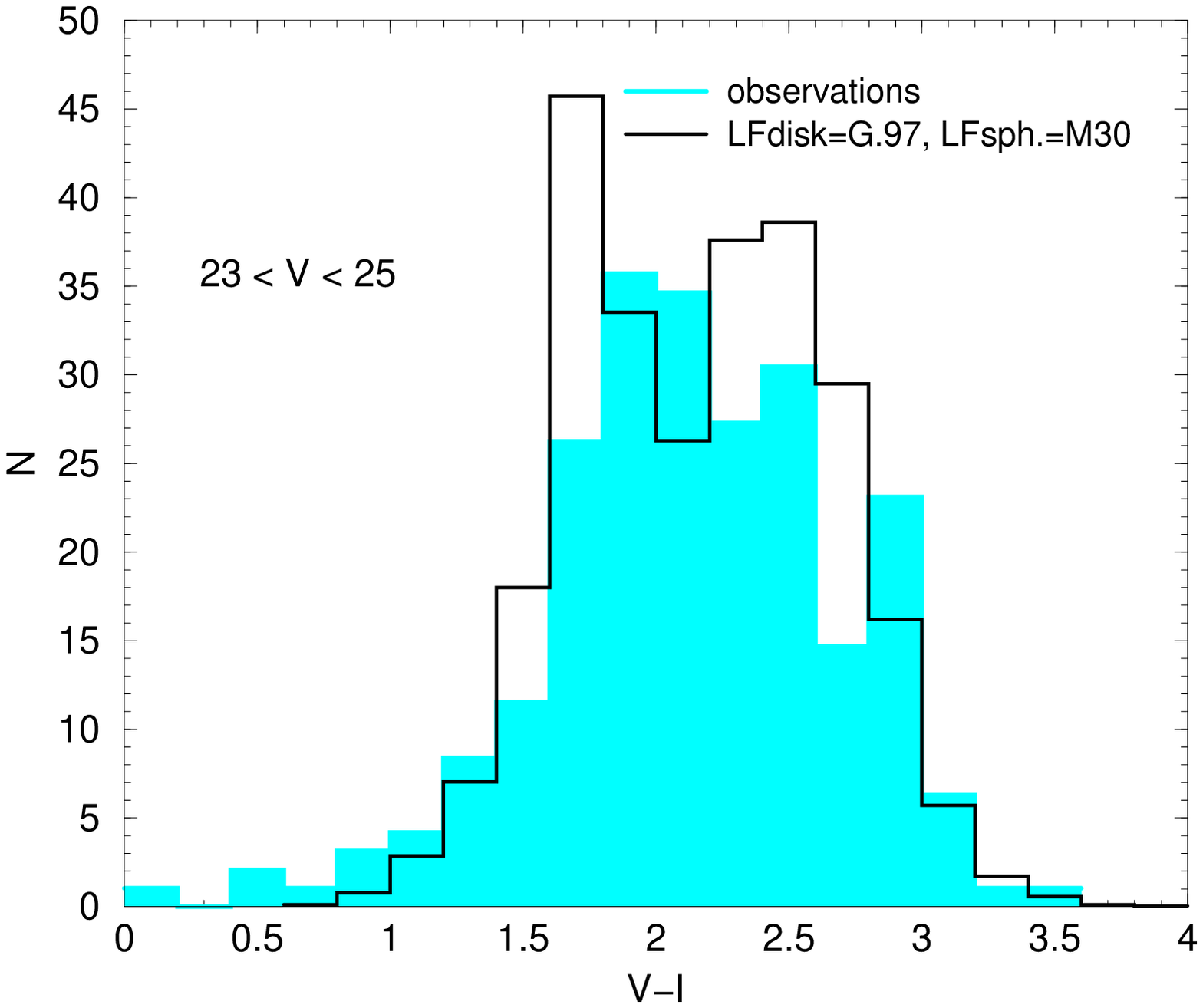}}
\caption{Theoretically predicted colour distributions for 23$<$V$<$25
for the models of Fig.8, as compared with observational data (shaded
region). Upper left panel: Wielen et al. (1983) disk LF + Dahn
(completed with M30) spheroid LF. Upper Righ Panel: Wielen et
al. disk LF + M30 LF for the spheroid. Lower Left Panel: Gould et
al. (1997) for the disk LF and Gould et al. (1998) for the spheroid
LF.  Lower Right Panel: Gould et al. (1997) for the disk LF and M30
for the spheroid LF. }
\end{figure}


Figure 8 shows that it is possible to find several combinations of LFs
for disk/spheroid which fit the observational magnitude distribution
in a better way.  However, due to the large uncertainties still
present in the determination of disk and spheroid LFs this seems to us
a meaningless numerical exercise.  Figure 9 shows the predicted color
distribution in the region 23$<$V$<$25 (the only affected by the
changes in the adopted LFs) for the combinations of LFs shown in
Fig.8. All the results are in fair agreement with observational color
distributions. Thus Fig.9 does not allows to discriminate among
different LFs.

\section{Models with thick disk}

In the previous sections we have shown that the two-component (disk +
spheroid) B\&S model, appears able to reach a satisfactory agreement
with the deep star counts provided by HST in the field of NGC 6397
only if LFs are larger than
expected on the basis of Gould et al. (1997,1998) indications. In this
section we will include a thick disk component to check if such a
conclusion is affected by the addition of a third component to the
Galactic structure.
 
In 1993 Gould et al. discussed photometric data
from the Hubble Space Telescope Snapshot Survey up to an average
apparent magnitude of V=21.4 to conclude that, mainly due to the lack
of information about the colors of the observed stars, it was not
possible to distinguish between the presence or the absence of the thick disk.
In 1996 Basilio et al. used I magnitude and ($V-I$) color star counts from
17 HST deep fields data to discriminate between the results of the
standard models by Bahcall \& Soneira (1980,1984) and by Gilmore \& Reid
(1983). They concluded that the two standard model predictions for the
magnitude counts was similar and in good agreement with 
observations. However they obtained a reasonable agreement between
model and data for colour distributions only with the introduction of
a thick disk component. Mendez \& Guzman (1998) by adopting a small
data sample with V up to $\approx$25 found an equally good agreement
between models with thick disk and observations for a thick disk with
a local normalization of 2\% and a scale height of 1300 pc (Reid \&
Majewski 1993) or a local normalization of 6\% and a scale height of
750 pc (Ojha et al. 1996). More in general, one finds that a large
variety of disk/thick disk parametrizations is available in literature
(see e.g. Reid \& Majewski 1993, Ojha et al. 1996).

 For our first test we adopted a consistent set of disk/thick
disk/spheroid parameters as given for the ``Besancon model'' by
Haywood, Robin \& Creze (HR\&C, 1997, see also Haywood, Robin \& Creze
1996, Ojha et al. 1996, Robin et al. 1996 for more details). The disk
/thick disk luminosity function adopted by the authors is from Wielen
et al. (1983) for M$_V$$>$4, from McCuskey (1966) for M$_V$ $<$-1 and
an average of both LFs in the interval -1$\leq$ M$_V$ $<$4.  The
chosen parameters are: a disk scale height of 240$\pm$50 pc, a thick
disk scale height of 760$\pm$50 pc, a disk scale lenght of
2500$\pm$300 pc, a thick disk scale lenght of 2500$\pm$800 pc, a thick
disk/disk local density ratio of 5.6$\pm$1\%, a spheroid/disk local
density ratio of 0.10\%, R$_{\circ}$=8.5 Kpc (see HR\&C). However
the ``Besancon'' model explicity allows for
time evolution of vertical scaleheight whereas the other models implicit
take into account this variation by assigning lower scaleheights to
early-type dwarfs. Thus here we will take a scale height of 90 pc for
the youngest MS stars and 240 pc for the oldest ones.

To explore the range of possible results, we made tests by adopting
either Wielen et al. (1983) or Gould et al. (1997) disk and thick disk
LFs for selected assumptions about the spheroid LF and by adopting for
the thick disk LF at bright magnitudes (M$_V$$<$4.5) the LF of 47 Tuc
(Hesser, Harris \& Vandenberg, 1987), assumed more representative of
high luminosity thick disk stars.  However numerical results show that
by replacing, for M$_V$$<$4.5, the 47 Tuc LF with the Wielen et al. LF
the differences in the resulting star counts, at least for our field,
are small.

Figure 10 shows the theoretical magnitude distributions when different
disk/thick disk and spheroid luminosity functions are used. The
contribution of the thick disk component is also shown.  The dashed
line represents a model with Wielen et al. (1983) disk/thick disk
luminosity function and Gould et al. (1998) spheroid LF at faint
magnitudes (M$_V$$>$ 8) implemented at brighter magnitudes by M30 LF
(Sandquist 1999). The solid line represents a model with Gould et
al. (1997) disk/thick disk LF and with Dahn et al. (1995) spheroid LF
at low luminosities completed with M30 for (M$_V$$<$ 8).  As expected,
one finds again that the distribution is affected by the change of the
faint LF only for M$_V$ $\gcu$ 24. However now one finds that the
three-component model with the parameters described above slightly
overestimates the counts up to (M$_V$$\approx$24). Comparison with
results in Fig.8 shows that predictions based on Gould et al. (1998) LF
appear in better agreement with faint star counts (see e.g.  the case
of Wielen et al. disk LF and Gould et al., 1998, spheroid LF).


\begin{figure}
\label{Mvthick}
\centerline{\epsfxsize= 8 cm \epsfbox{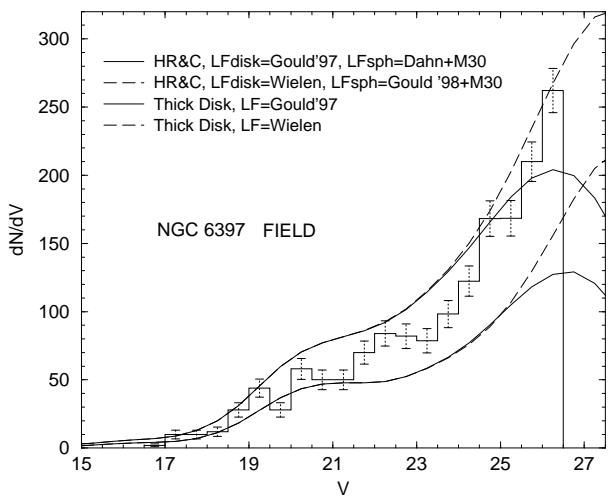}}
\caption{The histogram of the observed magnitudes (as in Fig. 2) as
compared with theoretical predictions by the B\&S model with a thick
disk component with parameters as in Haywood et al. (1997) (HR\&C). The
results for the two labelled different combinations of disk (thick
disk)/spheroid LFs are shown (see text). The contribution of the thick disk
component to the total counts is also shown.}
\end{figure}


\begin{figure}
\label{Oju}
\centerline{\epsfxsize= 8 cm \epsfbox{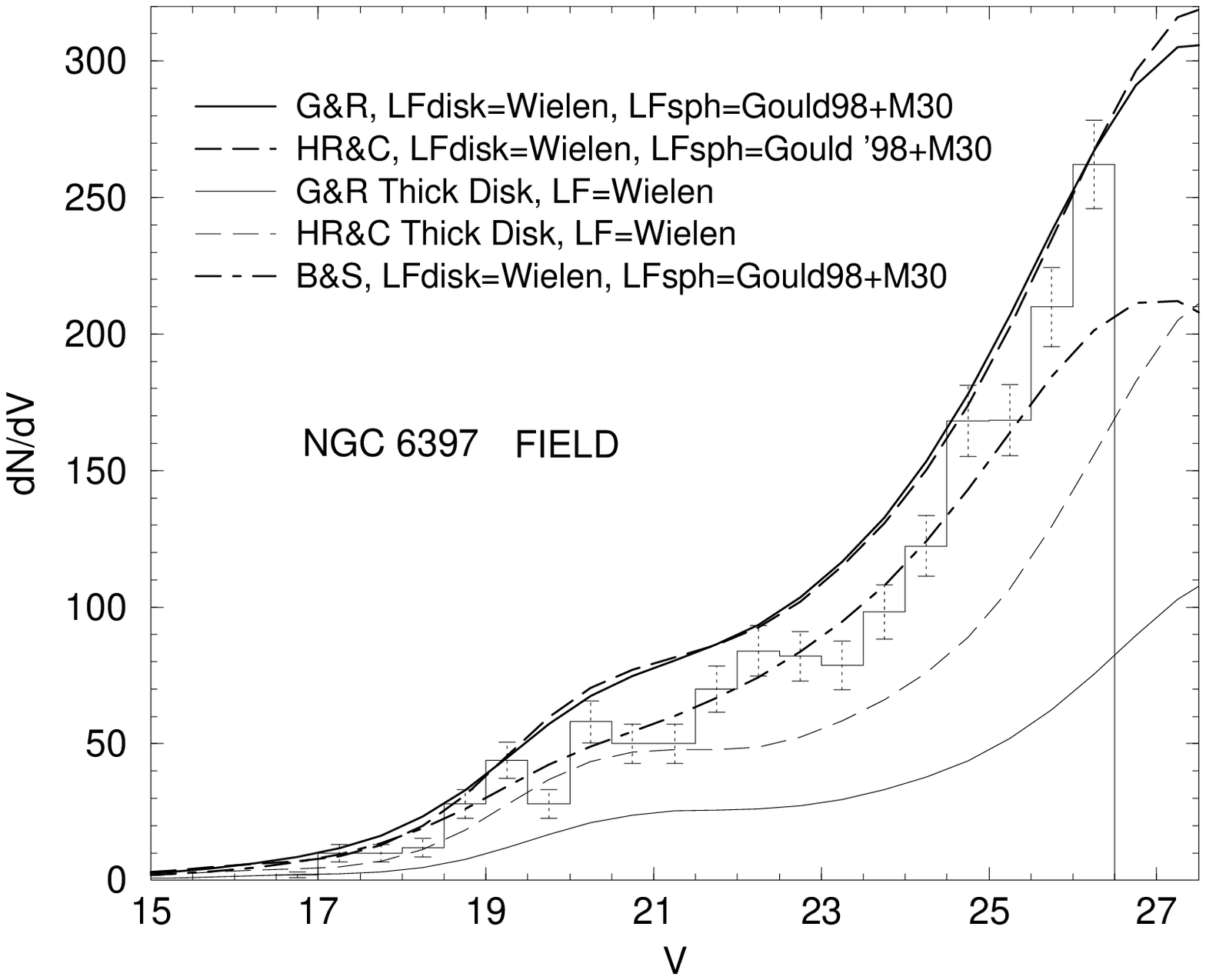}}
\centerline{\epsfxsize= 8 cm \epsfbox{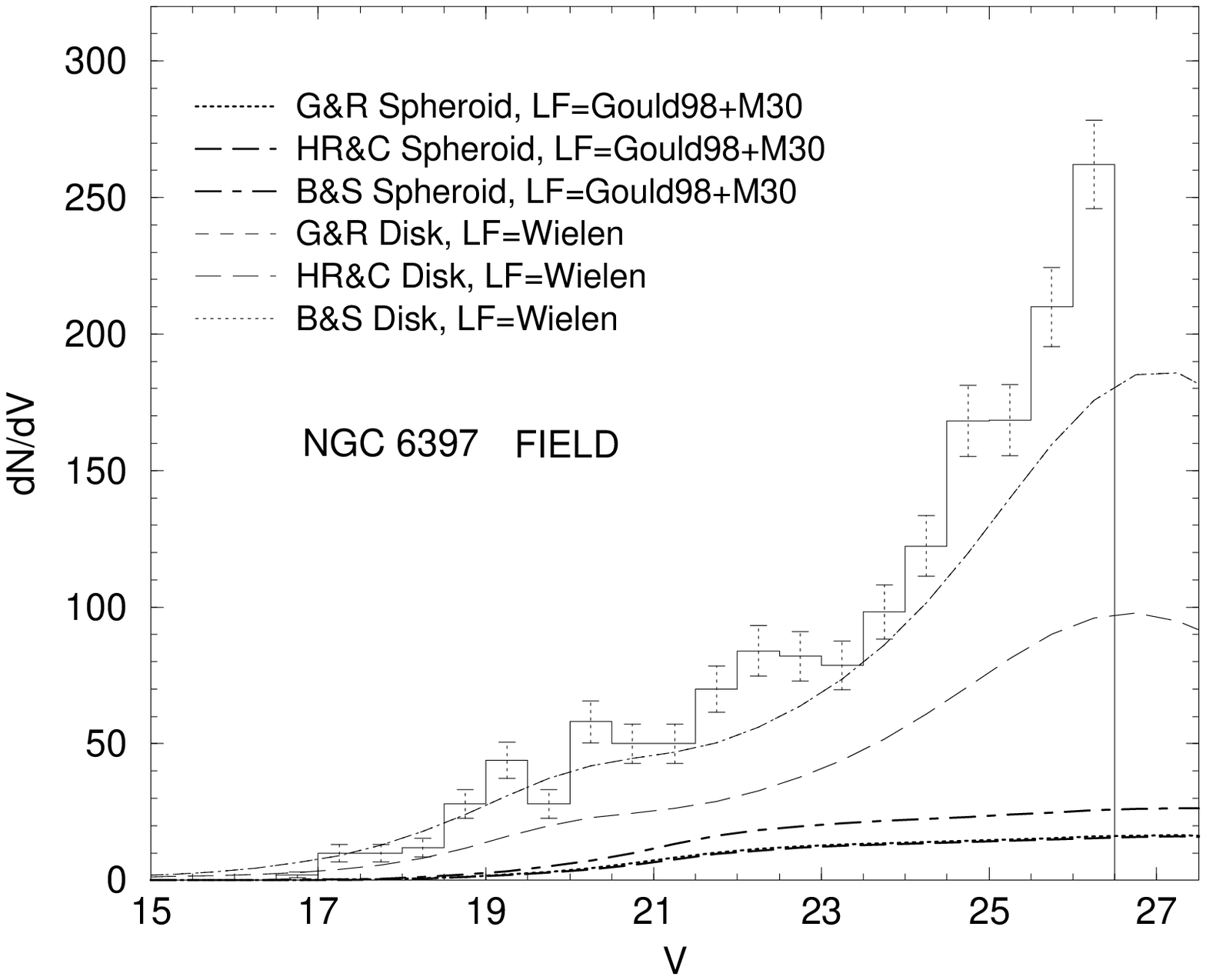}}
\caption{Upper panel: comparison between observational data and predicted 
magnitude distributions for a three-component model with Gilmore \& Reid (G\&R,
1983) parameters and Haywood et al. (HR\&C, 1997) parameters (see
text). The adopted LFs are Wielen et al. (1983) for the disk/thick
disk (completed in this last case with the LF of 47 Tuc) and Gould et
al. (1998), completed with M30 for M$_V$$<$ 8, for the spheroid. The
contribution to the total counts of the two thick disk components are
also shown. For comparison the two components B\&S model with the same disk/spheroid
LFs is also shown.  Lower panel: contribution to the star counts of
the disk/spheroid components for the three selected models.
}
\end{figure}


\begin{table*}
\caption{Parameters of the two components Bahcall-Soneira (B\&S)
model and of the models with thick disk by Haywood, Robin \& Creze
(HR\&C) and by Gilmore \& Reid (G\&R), see text for more details.}
\label{tab1}
\begin{center}
\begin{tabular}{ c l c c c}
\hline
\hline\\
{\bf Models}    &                    &{\bf B\&S}& {\bf G\&R}  & {\bf HR\&C}  \\
\hline
{\bf Disk}      &  scale height (pc) &  325    &  325     &  240$\pm$50 \\ 
	        &  scale lenght (pc) & 3500    & 3500     & 2500$\pm$300\\
{\bf Thick Disk}&  scale height (pc) & ----    & 1300     &  760$\pm$50 \\ 
	        &  scale lenght (pc) & ----    & 3500     & 2500$\pm$300\\
                &  thick disk/disk   & ----    & 2\%      & 5.6$\pm$1\% \\
                &  density ratio     &         &          &             \\
{\bf Spheroid } &  spheroid/disk     & 0.200\% & 0.125\%  & 0.100\%     \\  
                &  density ratio     &         &          &             \\
                &   axis ratio       &  0.8    &   0.8    &   0.8       \\
\hline 
\end{tabular}
\end{center}
\end{table*}

Before discussing such an evidence, as an alternative choice, let us
now explore the model by Gilmore \& Reid (1983) by adopting their most
frequently used parameters (G\&R, 1983, 1984, see also Colless et
al. 1991, Basilio et al. 1996).  The chosen parameters are: an disk
scale height of 325 pc, a thick disk scale height of 1300 pc, a thick
disk scale lenght of 3500 pc, a thick disk/disk local density ratio of
2\%, a spheroid/disk local density ratio of 0.125\% with, as in Reid
\& Majewski (1993), similar LFs for both disk and thick disk.  Reid \&
Majewski (1993) adopted for disk/thick disk luminosity function the
Wielen et al. (1983) LF for M$_V$$<$ 11 implemented at fainter
magnitudes with the photometric-parallax-based LF by Reid (1987) which
is not too much different from the Reid \& Gizis one, shown in Fig.6
and thus it is intermediate among the Wielen et al. (1983) and Gould
et al. (1997) one. For reader's convenience, table \ref{tab1} summarizes
the main disk/thick disk/spheroid parameters of the selected galactic models.


\begin{figure}
\label{Ojunew}
\centerline{\epsfxsize= 8 cm \epsfbox{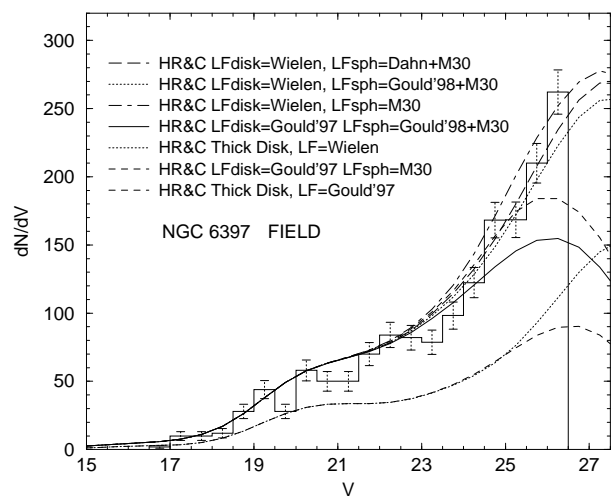}}
\caption{The histogram of the observed magnitudes (as in Fig. 2) as
compared with theoretical predictions by the B\&S model with a thick
disk component with parameters as in Haywood et al. (1997) (HR\&C) for
the labelled combinations of disk/thick disk/spheroid LFs at faint
magnitudes. The thick disk parameters are: scale height = 710 pc and
thick disk/disk local density ratio of 4.6\% (see text).  The
contribution of the thick disk component to the total counts is also
shown.}
\end{figure}


Fig.11 (upper panel) shows the comparison between the predicted
magnitude distribution for a model with the parameters of the
``Besancon model'' and the one with the Gilmore \& Reid model
parameters. The adopted luminosity functions are Wielen et al. (1983)
for disk and thick disk and Gould et al. (1998) + M30 for the
spheroid.  One sees that the predictions of the two models are
coincident in the whole region of interest (M$_V$ $\lcu 26.5$).  This
result confirms that different combinations of disk/thick
disk/spheroid parameters can give the same results for predicted star
counts (see e.g. the discussion in Mendez \& Guzman 1998) and thus
there is no an unique solution when the results of different models
are compared with observational magnitude and color counts. Mendez \&
Guzman (1998) concluded that to discriminate among different possible
parameters one should need kinematical data in addition to star counts
as done e.g. by Ojuka et al. (1996) who obtained the parameters
adopted in the ``Besancon'' model. Fig.11 also shows the contribution
of the thick disk components to the total counts. Note that while the
total counts are very similar for the two sets of parameters (G\&R and
HR\&C) the contribution of the thick disk is different, due to the
different normalizations, scale heights and scale lenghts. For
comparison the total star counts of the two
components B\&S model with the same disk/spheroid LFs is also
shown. As expected, due to the lack of the thick disk
component the total counts of the B\&S model are lower.
Fig.11 lower panel shows the contribution to the total counts of the upper panel 
of disk and spheroid components for the three
selected models. The B\&S disk overlaps the G\&R one because the two
models adopt the same disk parameters (see Table 1), while the HR\&C
model adopts a lower scale height for the disk and thus its thick disk
contribution is larger (see upper panel). The spheroid contribution of
the B\&S model is larger with respect to the ones of the models with
the thick disk (see Table 1).

From Figs. 10 and 11 one should conclude that, for the field of NGC
6397, both the adopted three-component models overestimate star counts
at bright magnitudes.  However, we remind that Robin et al. (1996)
evaluate for the thick disk density and scale height an uncertainty
of, respectively, $\approx$20\% and $\approx$7\%. Within this
uncertainty the thick disk density and scale height for the
``Besancon'' model can thus be decreased down to 4.6\% and 710 pc,
respectively. The results of such a decrease are shown in Fig.12 for
various combinations of disk/thick disk/spheroid LFs at faint
magnitudes.


\begin{figure}
\label{Ojunew}
\centerline{\epsfxsize= 8 cm \epsfbox{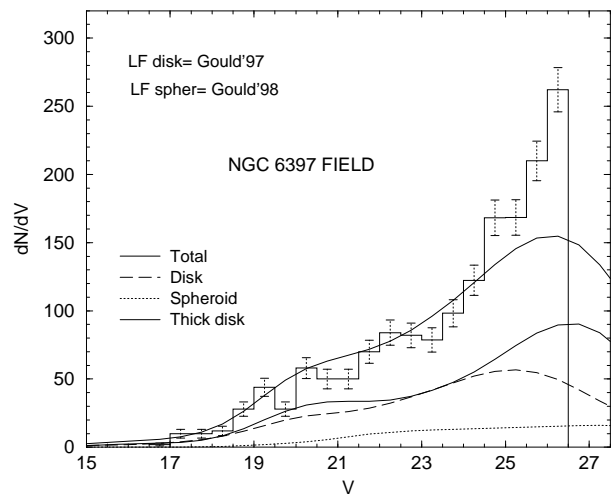}}
\caption{The histogram of the observed magnitudes (as in Fig. 2)
compared with theoretical predictions by the B\&S model with a thick
disk component with parameters as in Haywood et al. (1997) (HR\&C).
The adopted luminosity functions are Gould et al. (1997) (completed
for M$_V$$<$ 9 with Wielen et al. 1983) for the disk/thick disk and Gould
et al. 1998 (completed with M30) for the spheroid.  The contribution
to the total counts of the three components is shown.  The thick disk
scale height is 710 pc and the thick disk/disk local density ratio
is 4.6\% (see text).
}
\end{figure}


One finds that just by tuning the thick disk parameters
within their uncertainties it is possible to obtain a good fit in the
high luminosity region of the magnitude distribution, while 
the fit at faint magnitudes depends on  the choice of the
combination of disk/thick disk/spheroid faint LFs. 
If relying on such a model, one finds again that the combination of Gould et
al. (1997) disk/thick disk LF and Gould et al. (1998) halo LF (completed with
M30 LF) appears in disagreement with observations (Fig.13), whereas 
the separate adoption of Gould et al. (1997) LF
for the disk/thick disk or the Gould et al. (1998) LF for the halo
cannot be excluded.


\begin{figure}
\label{colOju}
\centerline{\epsfxsize= 4 cm \epsfbox{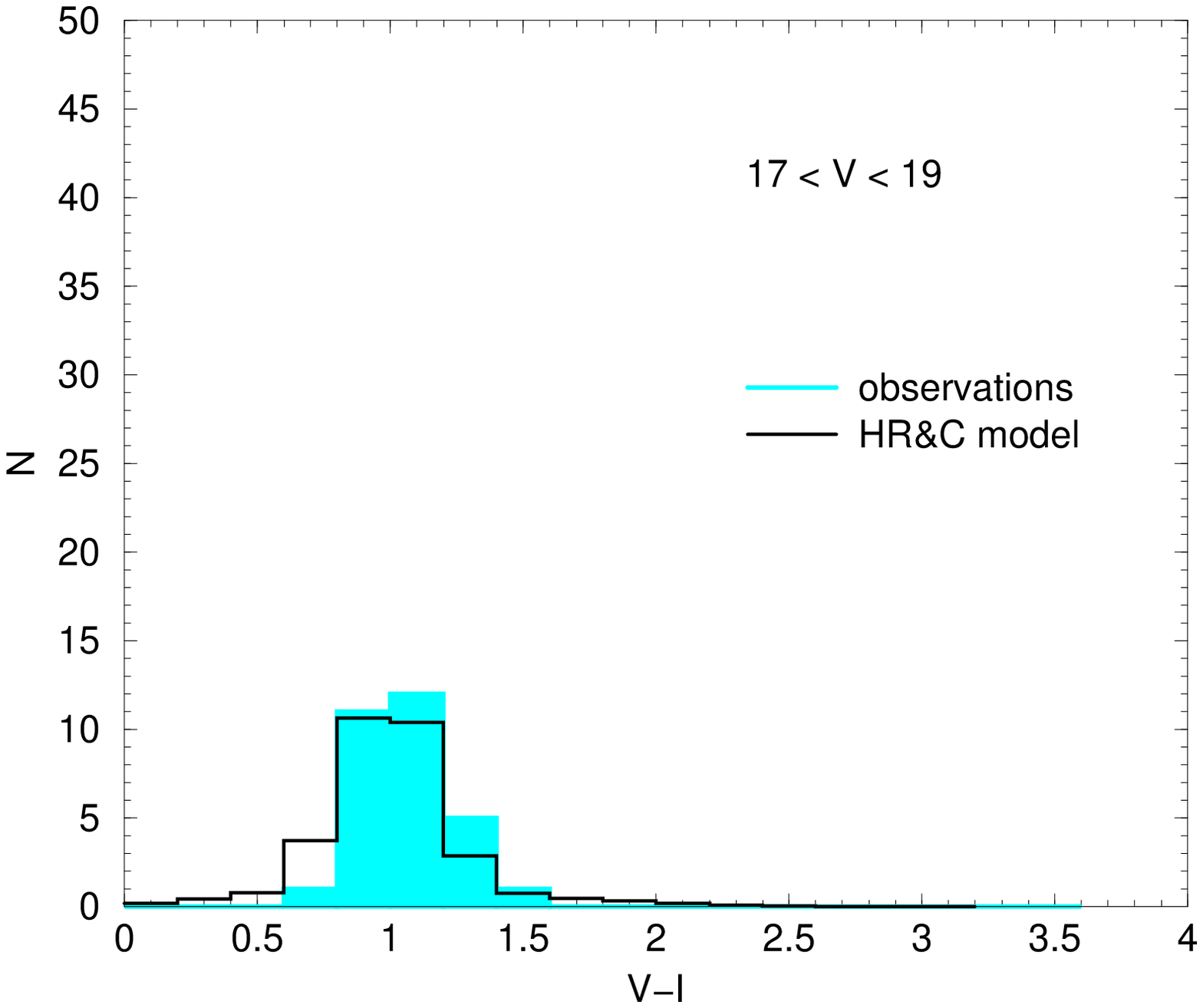 }
\epsfxsize= 4 cm \epsfbox{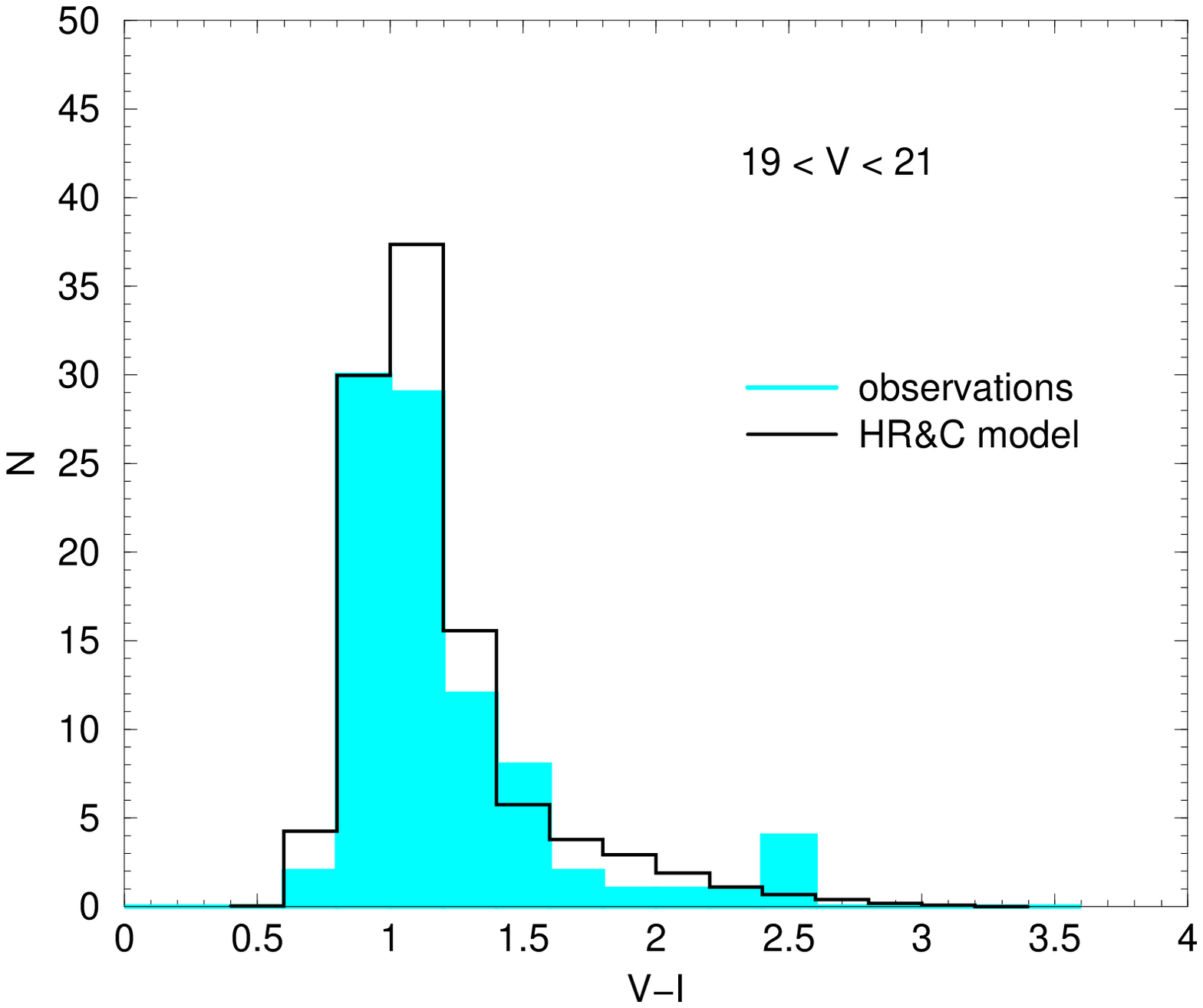}} 
\centerline{\epsfxsize= 4 cm \epsfbox{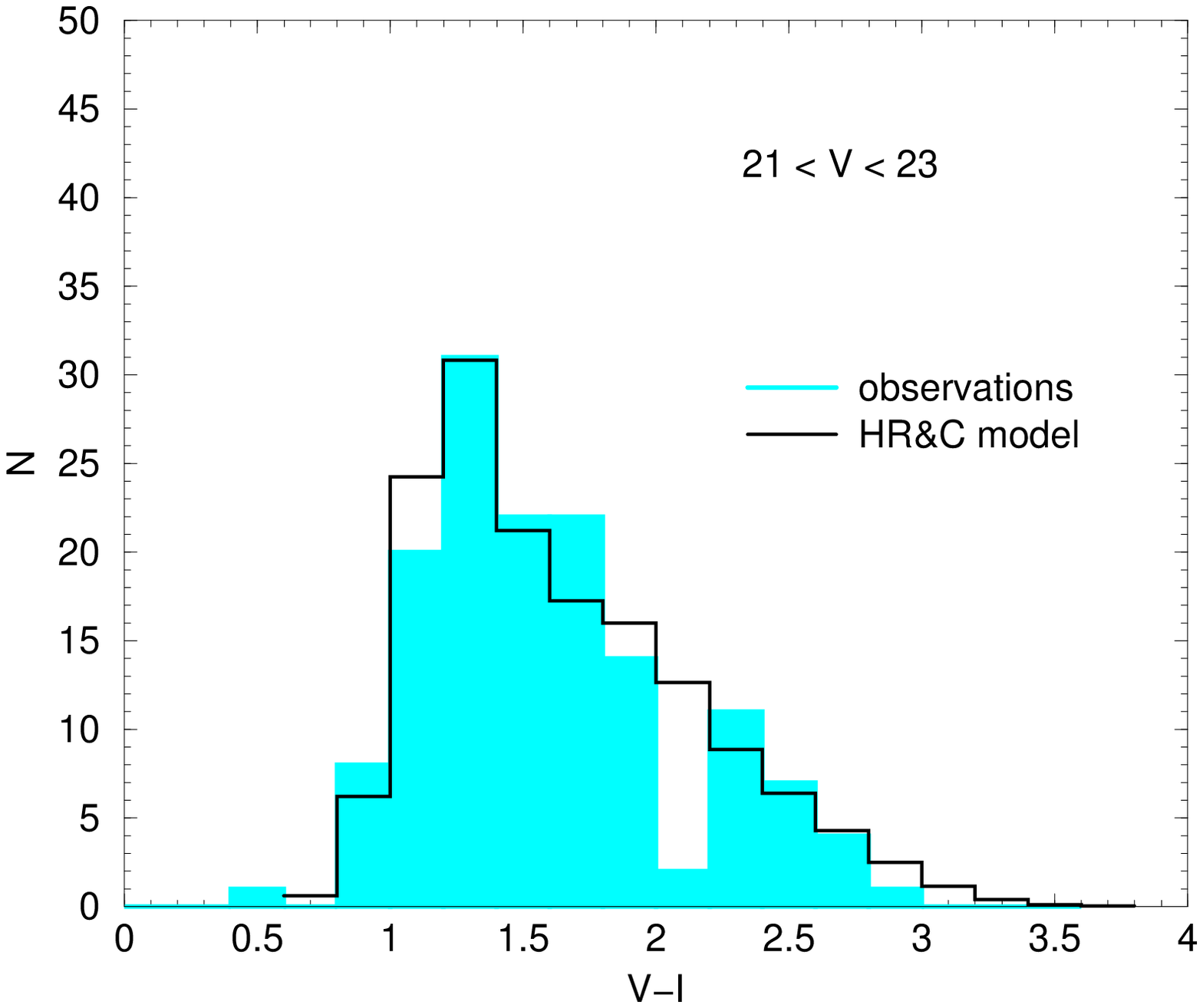}
\epsfxsize= 4 cm \epsfbox{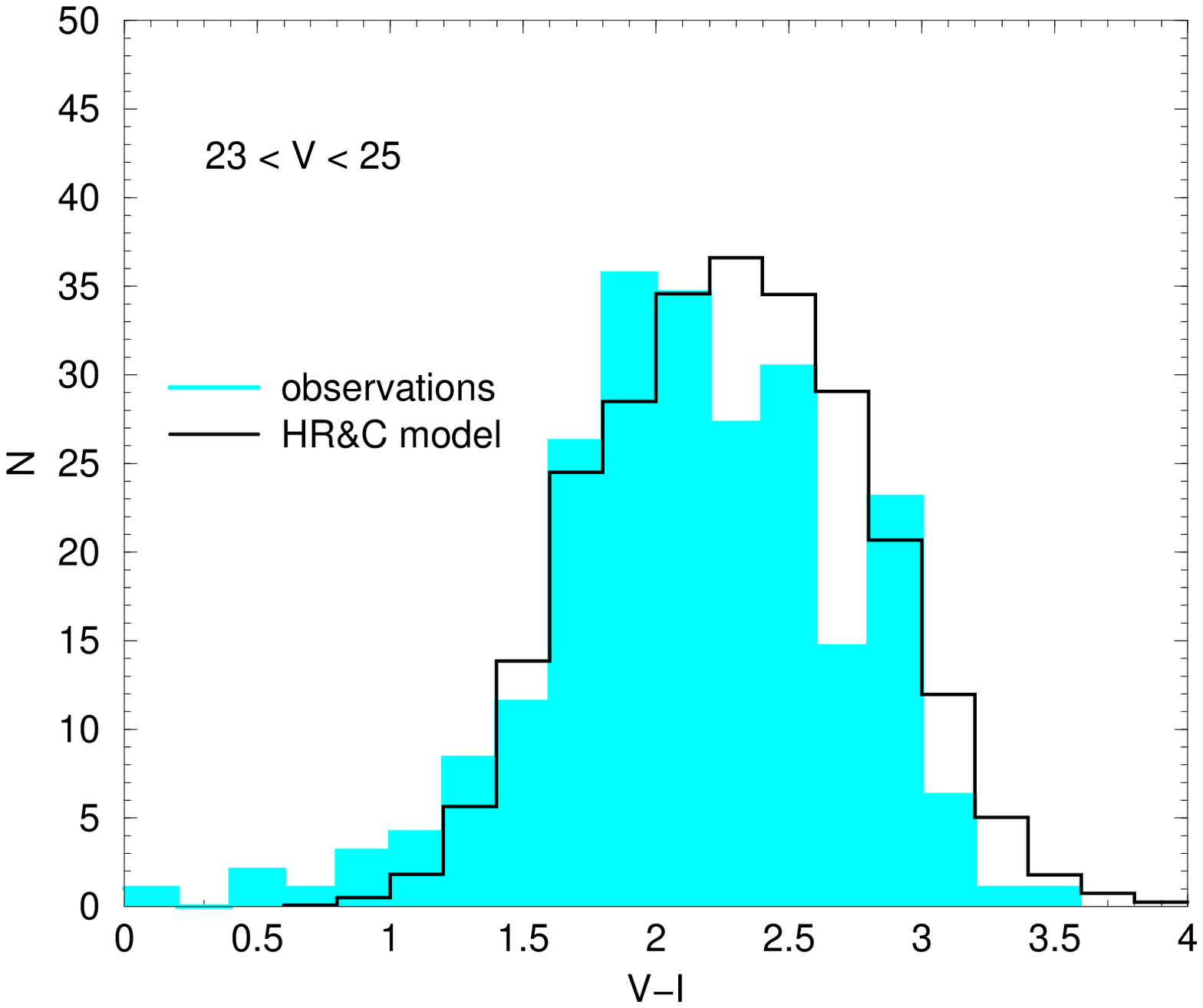}} 
\caption{Comparison between the observed (shaded region) and predicted
distribution of $V-I$ colours in the labelled intervals of magnitude
for the three-component model with Haywood et al. parameters. The
thick disk scale height and density are respectively 710 pc and 4.6\%
of the local disk (see text).  The adopted LFs are Wielen et
al. (1983) for disk and thick disk and Dahn et al. (1995), completed at bright
magnitudes with the LF of M30, for the spheroid component.}
\end{figure}

\begin{figure}
\label{contributi}
\centerline{\epsfxsize= 4 cm \epsfbox{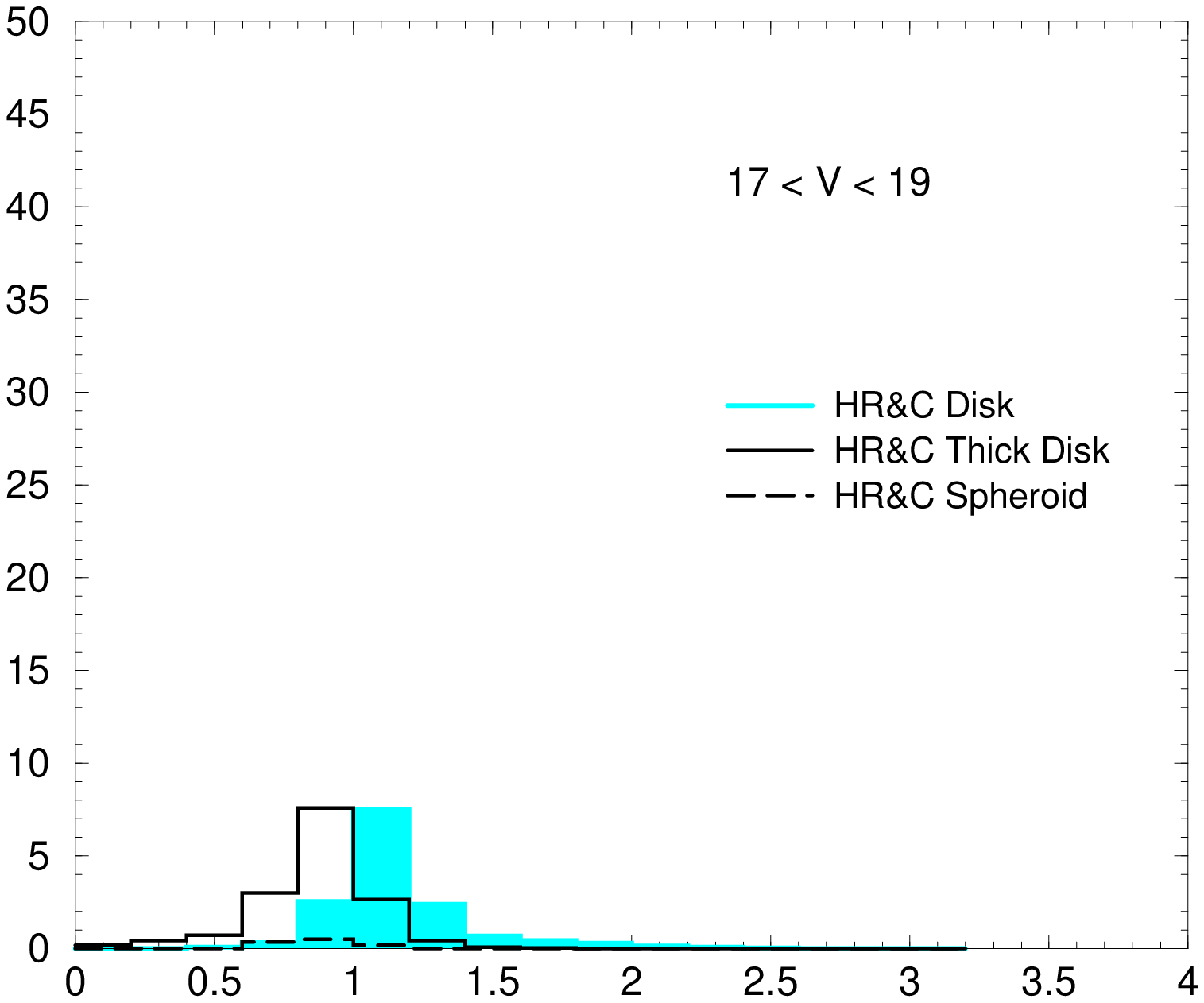}
\epsfxsize= 4 cm \epsfbox{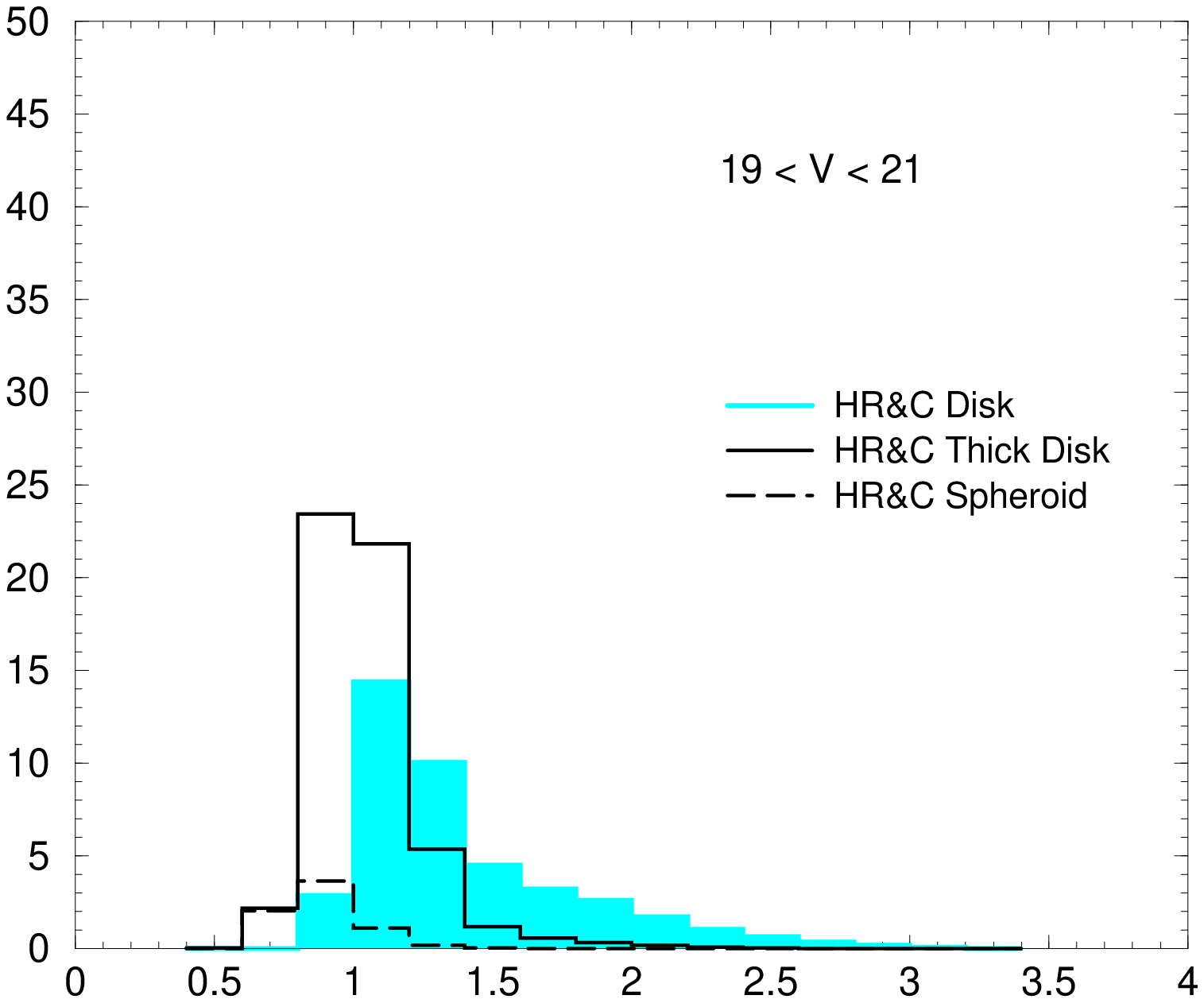}} 
\centerline{\epsfxsize= 4 cm \epsfbox{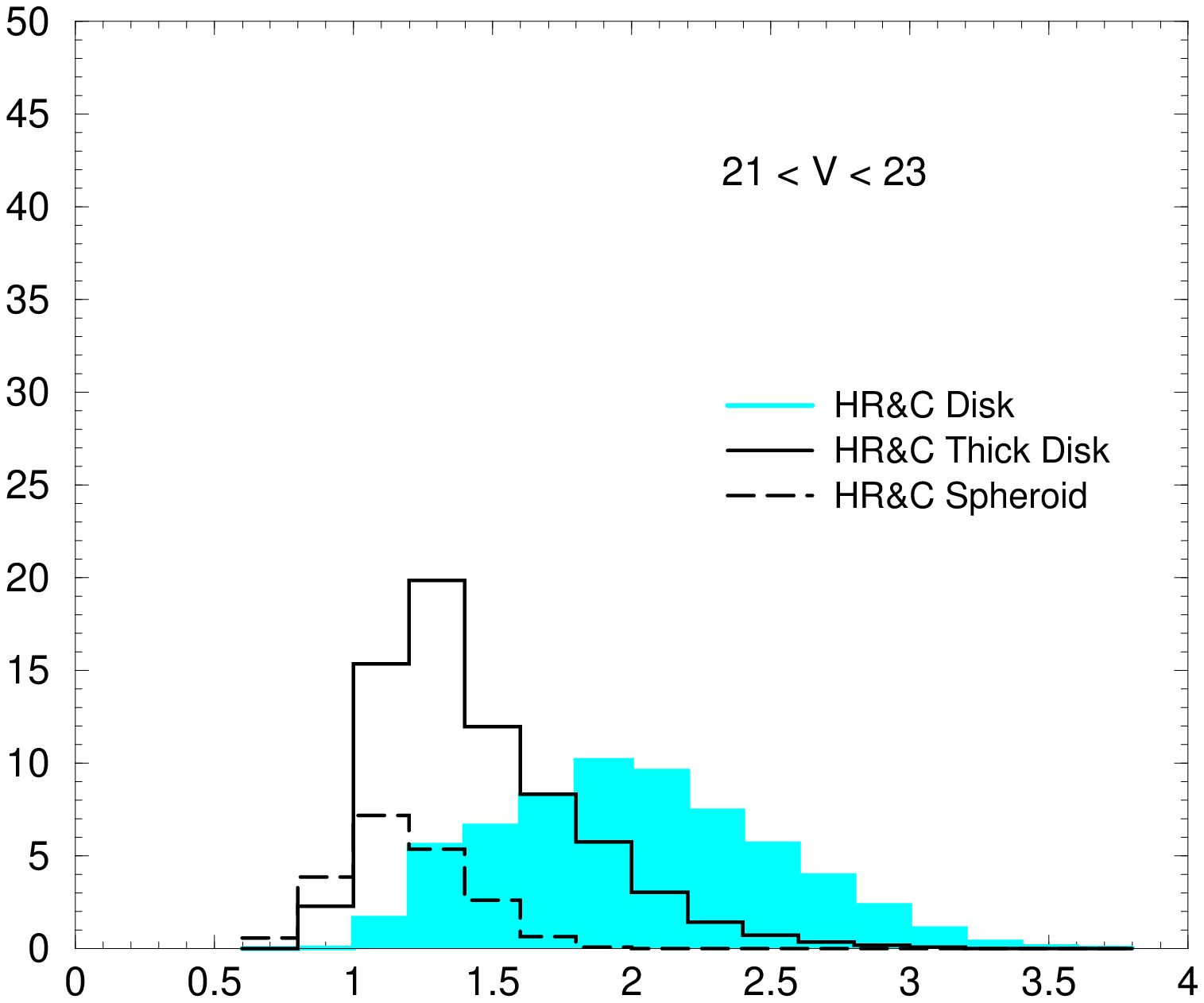}
\epsfxsize= 4 cm \epsfbox{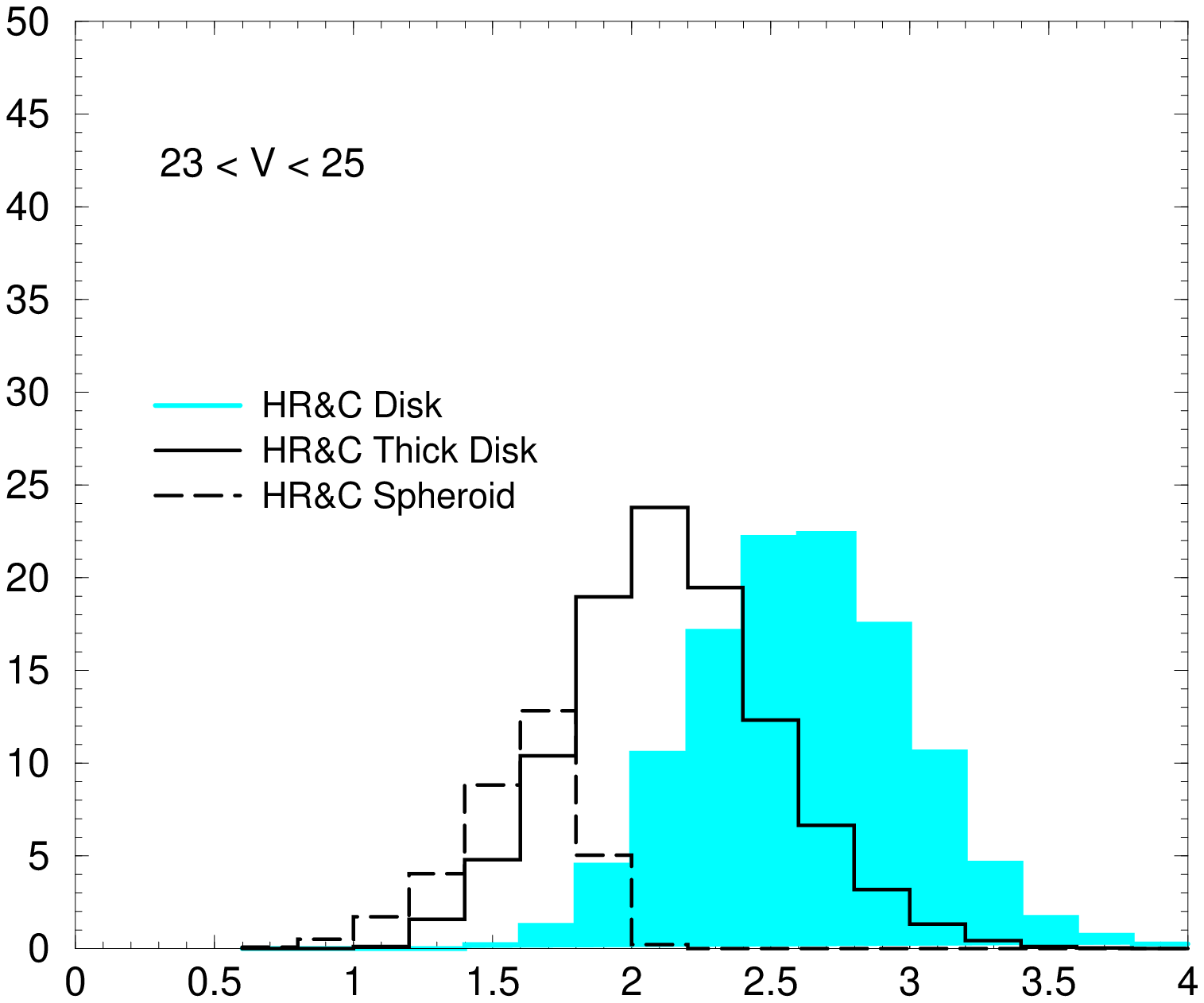}} 
\caption{Contributions of the three different populations for the colour-number plots
of Fig.14.}
\end{figure}

Fig. 14 shows the predicted ($V-I$) colour distribution in the labelled
intervals of magnitude with the `Besancon'' parameters tuned as in
Fig. 12 with the Wielen et al. LF for the disk and Dahn et al. 1995 (+
M30) LF for the spheroid (heavy long dashed line in Fig.12).
Following Gilmore (1984) (see also Reid \& Majewski 1993) we used the
CMD of the metal rich globular cluster 47 Tucanae to derive the
color-magnitude relation for evolved thick disk stars. For MS thick
disk stars we used theoretical predictions with Z=0.006 (Cassisi et
al. 2000) which, as already discussed, have shown to be in very good
agreement with observational data.  In all cases E(B-V)=0.18.  One
sees that the agreement is reasonable in all the magnitude intervals.
Figure 15 shows the contribution of disk, thick disk and spheroid    
to the colour-number plots of Fig.14. Due to the HR\&C choice of
normalizations, scale heights and scale lenghts, the thick disk contribution 
in the selected intervals of magnitude is slightly larger than the disk contribution
(see  Fig. 11).  Moreover the thick disk colors are bluer than the disk ones due the lower
metallicity (Z=0.006) of the thick disk.


\begin{figure}
\label{colOju2}
\centerline{\epsfxsize= 4 cm \epsfbox{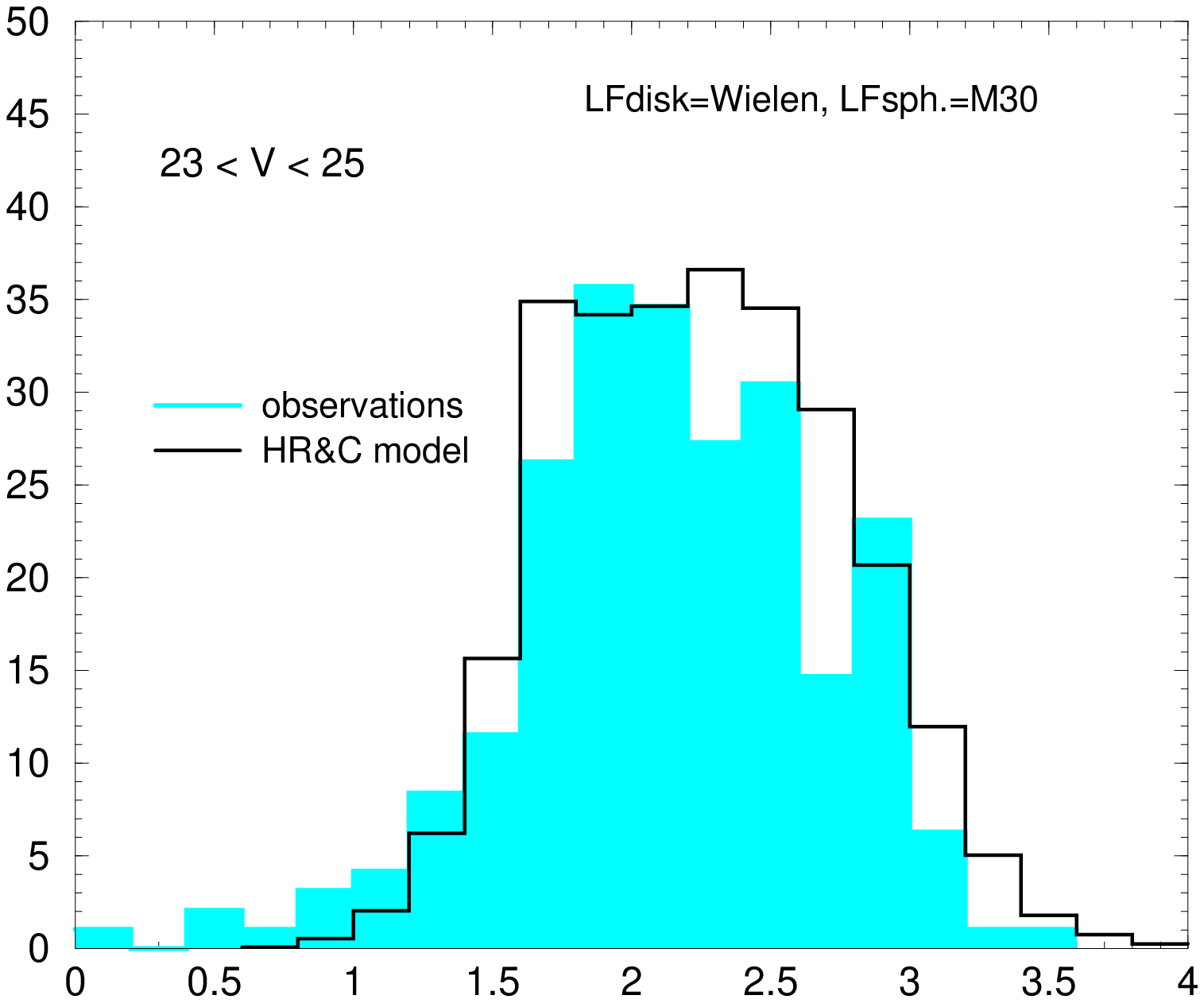}
\epsfxsize= 4 cm \epsfbox{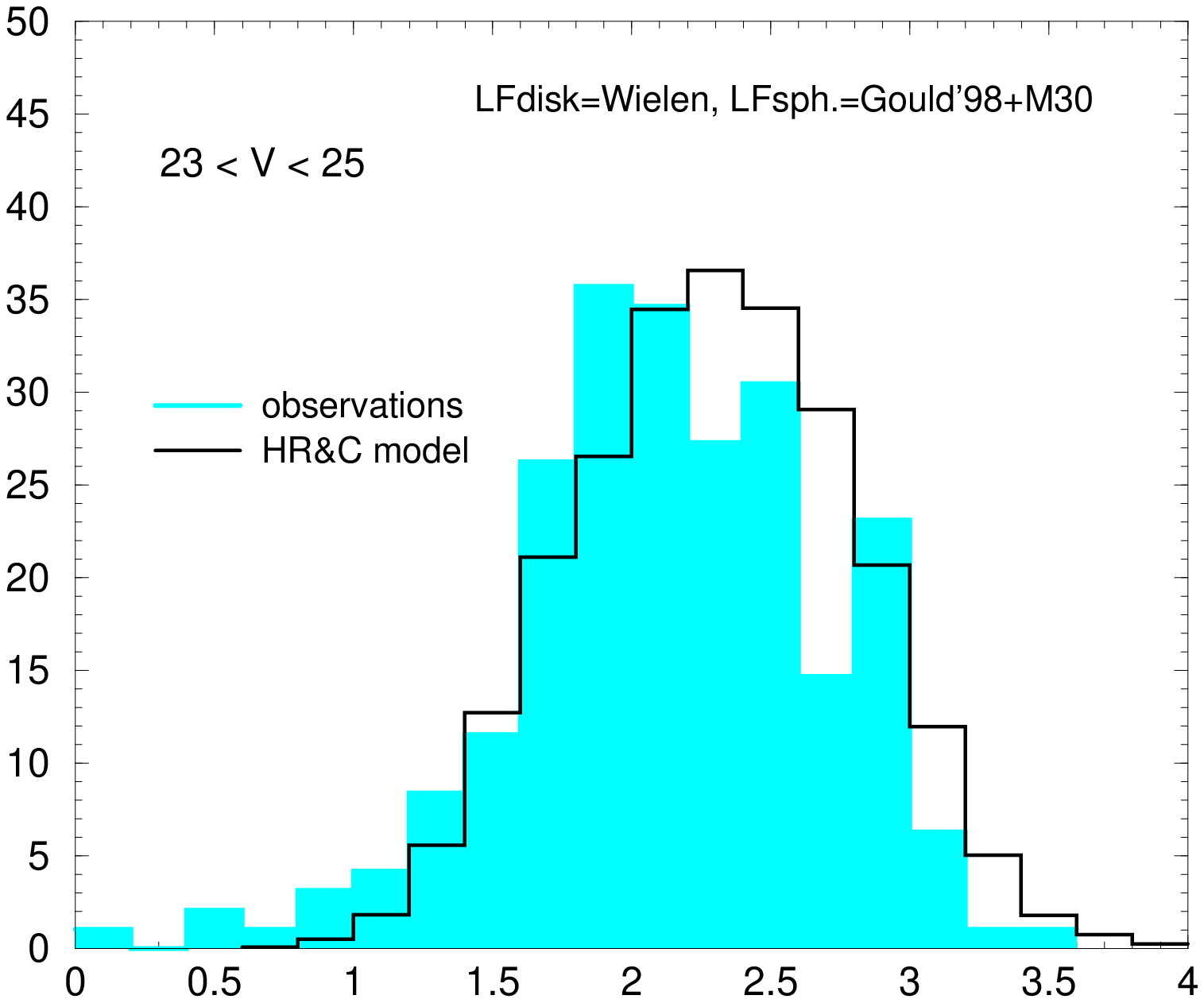}} 
\centerline{\epsfxsize= 4 cm \epsfbox{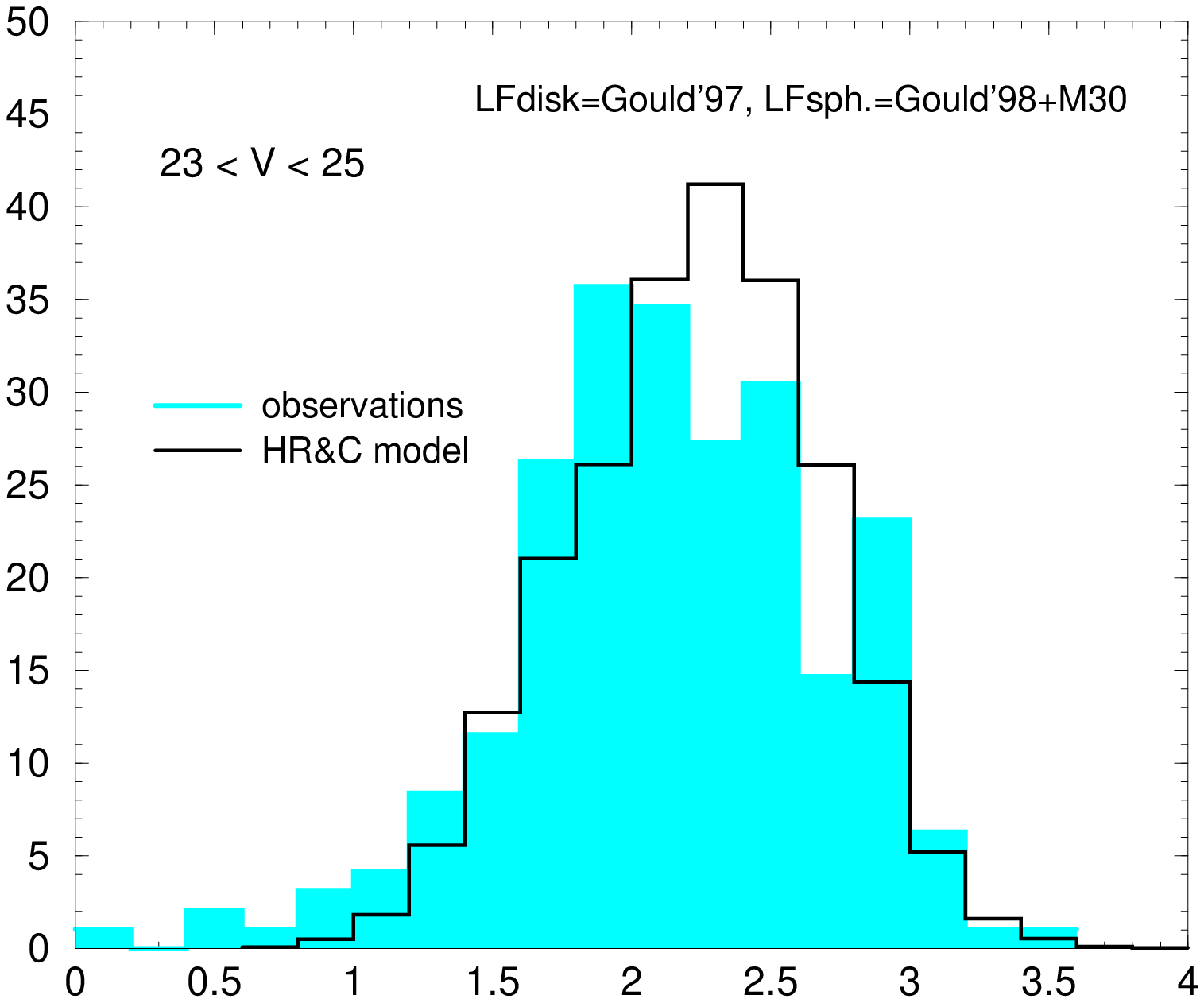}
\epsfxsize= 4 cm \epsfbox{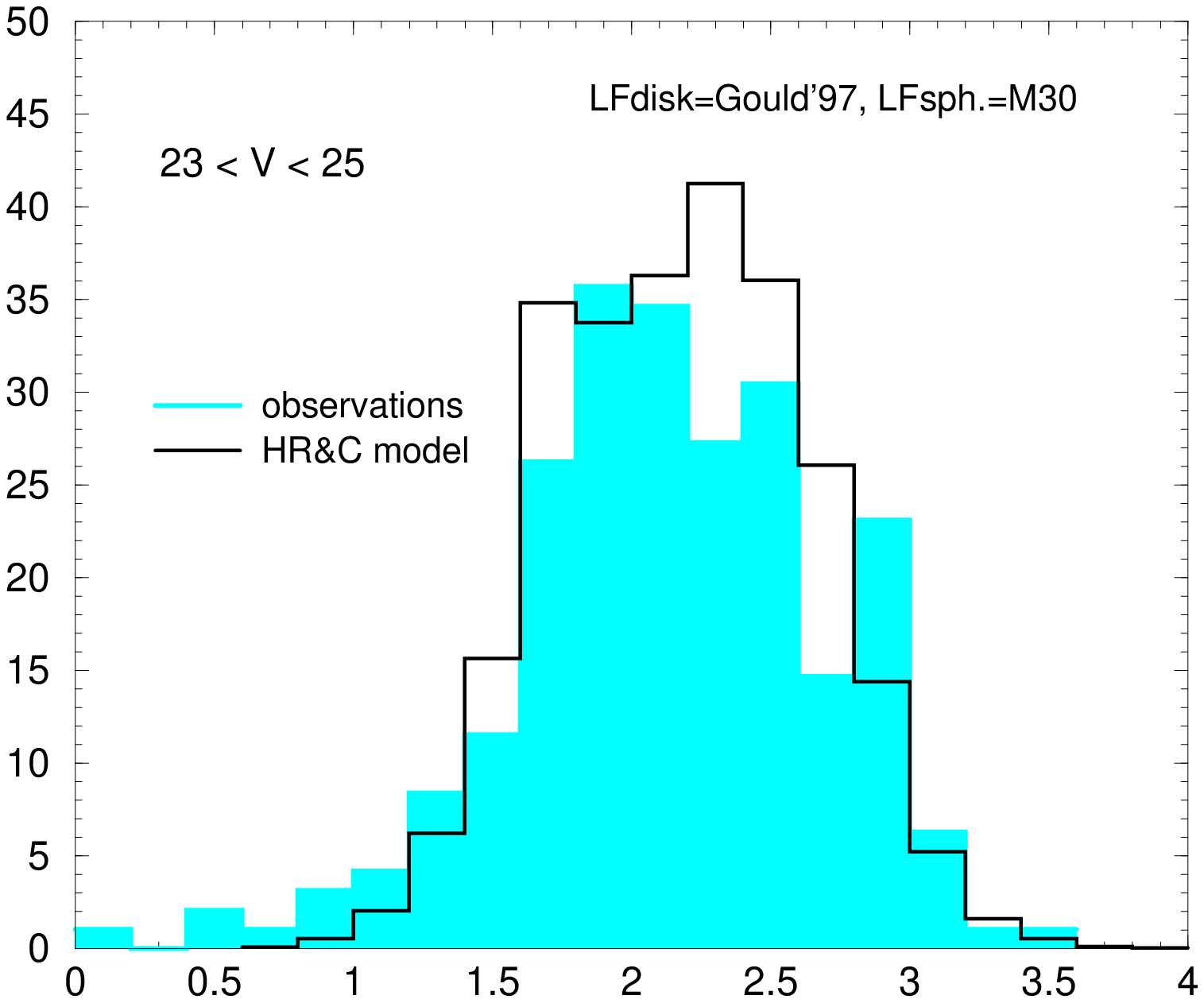}} 
\caption{Comparison between the observed (shaded region) and predicted
distribution of $(V-I)$ colours for 23$<$V$<$25 for the three component
model with Haywood et al. parameters and for different combinations
of disk/thick disk/spheroid LFs (as labelled). The thick disk scale height and
density are respectively 710 pc and 4.6\% of the local disk (see
text).}
\end{figure}


\begin{figure}
\label{Ojunew}
\centerline{\epsfxsize= 8 cm \epsfbox{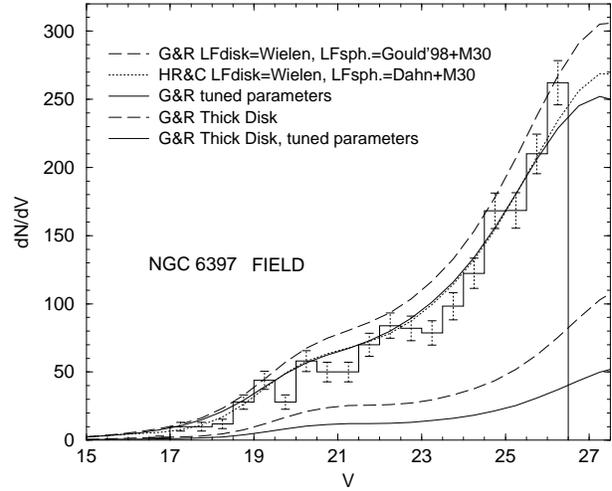}}
\caption{Comparison among theoretical magnitude distribution for the
Gilmore \& Reid model with original (heavy dashed line) and tuned
parameters (heavy solid line, see text).  The adopted LFs are Wielen
et al. for the disk/thick disk and Gould et al. 98 (+M30) for the
spheroid.  For comparison the model with the HR\&C parameters tuned as
in Fig. 12 is also shown (dotted line); the adopted LFs are Wielen et
al. for the disk/thick disk and Dahn et al.(+M30) for the spheroid.  The
thick disk contribution to the total counts is also shown.  }
\end{figure}


Figure 16 compares observed and predicted colour counts for
23$<$V$<$25 for different assumptions about the faint LFs for
disk/thick disk and spheroid (see Fig.12 for the corresponding
magnitude distributions).  As in the two-component model, all the LF
combinations taken into account (even Gould et al. 1997 - Gould et
al. 1998) show an acceptable fit to observational data; thus the
color distributions does not allow to discriminate among different LFs
at faint magnitudes.

For the Gilmore \& Reid model too the thick disk parameters can be
tuned to reconcile the model with observations.  In this case one
expects that due to the lower contribution of the thick disk to the
counts (see Fig.11) the effect of varying the thick disk parameters
will be lower. In fact, taking as representative the estimate of
errors by Robin et al. (1996), to obtain a good agreement with
observational data, one needs to reduce the thick disk scale height
and normalization by 2$\sigma$ (scale height=1100 pc,
normalization=1.25\%).  Figure 17 shows the magnitude distribution for
the model with original Gilmore \& Reid (1983) parameters, (with
Wielen et al. disk/thick disk LF and Gould et al., 1998, +M30 spheroid
LF) together with the corresponding results when the thick disk
parameters are lowered by 2$\sigma$. For comparison the model with the
HR\&C parameters tuned as in Fig. 12 is also provided.  With this
modifications the results of the G\&R model appear as good as the
HR\&C one. We do not show the color distributions for the ``modified''
G\&R model because as disclosed by Fig.17, the results are practically
coincident with those of Fig.14 and Fig.16.

\section{Conclusion}

In this paper we have discussed star counts in the field of the
globular NGC 6397. As expected, we found that the predicted
distribution of stars fainter than M$_V$$\approx$24 is critically
dependent on the assumptions about the LF of faint MS stars.
According to the discussion given in the previous sections one finds
that neither the two-component B\&S model not the more sophisticated
three-component models by HR\&C and G\&R can satisfactory fit star
counts in the field of NGC 6397, when assuming for disk and spheroid
the low LF suggested by Gould et al. (1997), Gould et al. (1998),
respectively.  A satisfactory fit of the data can be achieved, either
for a two-component or for a three component model, with other
suitable combinations of LFs suggested in the current literature,
provided that density and scale height of the two discussed
three-component models are decreased within the predicted existing
uncertainties.  Nor the color distributions appear able to
discriminate among the various possible solutions.
However, we regard the previous conclusions as suggestions to
be further tested in deep surveys at different Galactic locations.  In
this context, and before closing the paper, one has to notice that
from the comparison of Figs.\ 2 and 6 one derives the tantalizing
evidence that going deeper than $V$=26.5 by a couple of magnitudes one
would derive much more stringent and precise constraints on the LF of
faint stars and, in turn, on the mean metallicity of the spheroid.
Such an evidence should be taken into account in future researches on this
matter.

\section*{Acknowledgments}

It's a pleasure to thank I. King for useful comments and for a careful
reading of an early version of the manuscript. We warmly thank
J.N. Bahcall for his advice and for making available to the scientific
community the Bahcall-Soneira Galactic model. GP acknowledges partial
support by the Agenzia Spaziale Italiana (ASI) and by the Ministero
della Ricerca Scientifica e Tecnologica (MURST) under the program
``Stellar Dynamics and Stellar Evolution in Globular Clusters''.  VC
and SDI thank the partial support by MURST within the ``Stellar
Evolution'' project.

\label{lastpage}

\end{document}